\documentclass{article}
\usepackage{pifont}
\usepackage{amsmath}
\usepackage{paralist}
\usepackage{graphicx}
\usepackage{float}
\usepackage{appendix}
\renewcommand{\appendixname}{Appendix~\Alph{section}}
 \usepackage[square, comma, sort&compress, numbers]{natbib}

\usepackage{amsmath}

\usepackage{parskip}
\usepackage{graphicx}
\usepackage{subfigure} 
\usepackage{array}
\usepackage{booktabs} 
\usepackage{multirow}
\usepackage{natbib}
\usepackage{setspace}
\usepackage[colorlinks,
            linkcolor=blue,
            anchorcolor=blue,
            citecolor=blue]{hyperref}
\usepackage{caption}
\DeclareCaptionLabelFormat{cont}{#1~#2\alph{ContinuedFloat}}
\newcommand{\tabincell}[2]{\begin{tabular}{@{}#1@{}}#2\end{tabular}}

\usepackage[left=1.3in,top=0.6in,right=1.3in,bottom=0.8in]{geometry} 

\title{\bf  Principal Component Analysis and Factor Analysis for Feature Selection in Credit Rating}
\author{ \bf Shenghuan Yang \\ \small  Jiangxi University of\\ \small Finance and Economics 
\and \bf lonut Florescu\\ \small Stevens Institute of\\  \small Technology \and \bf Md Tariqul Islam\\ \small  Syracuse University
}
\date{}

\usepackage{natbib}
\usepackage{graphicx}
\usepackage{xcolor}
\usepackage{xpatch}
\begin{document}

\maketitle

\begin{abstract}
 The credit rating is an evaluation of a company's credit risk that values the ability to pay back the debt and predict the likelihood of the debtor defaulting. There are various features influencing credit rating. Therefore, it is essential to select substantive features to explore the main reason for credit rating change. To address this issue, this paper exploited Principal Component Analysis and Factor Analysis as feature selection algorithms to select important features, summarized the similar features together, and obtained a minimum set of features for four sectors, Financial Sector, Energy Sector, Health Care Sector, Consumer Discretionary Sector. This paper used two data sets, Financial Ratio and Balance Sheet, with two mappings, Detailed Mapping, and Coarse Mapping, converting the target variable(credit rating) into categorical variable. To test the accuracy of credit rating prediction, Random Forest Classifier was used to test and train feature sets. The results showed that the accuracy of Financial Ratio feature sets was higher than that of Balance Sheet feature sets. In addition, Factor Analysis can reduce the number of features significantly to obtain almost the same accuracy that can decrease dramatically the time spent on analyzing data; we also summarized seven dominant factors and ten dominant factors affecting credit rating change in Financial Ratio and Balance Sheet  by utilizing Factor Analysis, respectively, which can explain the reason of credit rating change better.

\end{abstract}
\section{Introduction}
As financial markets become more complex and borrowers become more diverse, investors and regulators increasingly rely on professional and authoritative credit rating agencies' opinions. Credit rating is an assessment of a potential debtor (individual, enterprise, company, or government), predicting its ability to repay debt and an implicit prediction of the possibility of default by the debtor. Credit rating refers to assessing qualitative and quantitative information of potential debtors by credit rating agencies, including information provided by debtors and other non-public information obtained by analysts of credit rating agencies. Credit rating evaluation is a complex process, which considers many factors to assign a corresponding level to the enterprise. Therefore, it is necessary to hire domain experts to conduct evaluations, but such assessments are expensive.  At present, scholars mainly use three methods to study corporate credit: statistical methods(Regression Analysis, Discriminant analysis, Bayesian analysis){\color{blue}{\cite{{ref111}}}}-{\color{blue}{\cite{{ref333}}}}, non-statistical methods(Random Forest, Support Vector Machine,  Neural Network){\color{blue}{\cite{{ref44}}}}-{\color{blue}{\cite{{ref66}}}}, and combined methods{\color{blue}{\cite{{ref77}}}}. Non-statistical methods are the most common use, and people usually use Random Forest to select important features that significantly impact the credit rating, then train and test the data set to predict the accuracy of corporate credit rating. However, in our studies, we want to exploit other methods not usually used to analyze the credit rating. Therefore we utilized Principal Component Analysis and Factor Analysis to select important features, trained and tested them 100 times by using Random Forest Classifier, recorded the average prediction accuracy of credit rating. Besides, we also classify the features to reduce dimension for our high dimensional data and find out the latent factors that create a commonality.

\textbf{Objectives.} This work aims to select important features via Principal Component Analysis and Factor Analysis, and sort the top 20 features as the feature data sets, respectively. Simultaneously, we also use a machine learning model random forest to train and test the data sets and compare which method can better choose the features that impact the credit rating. Finally, we conclude the main causes that contribute to the corporate credit change.

\textbf{Contributions.} Following are the major contributions of this paper: 

\begin{itemize}
    \item We compared the Principal Component Analysis and Factor Analysis, which both identify patterns in the correlations between variables. We optimized  the  sorting method of these two techniques, and selected the top 20 important features as feature data sets.Then we trained and tested data sets to make prediction by utilizing Random Forest. Then we found that Factor Analysis needs fewer features than Principal Component Analysis to achieve the same accuracy(above 95\%), which reduces the time spent on data analysis to improve computational efficiency.
    \item We summarized similar features via Factor Analysis to determine the latent factors that explain one public aspect, and better illustrate the reason for credit rating change. For the data sets of  Financial Ratio and Balance Sheet, there were seven factors and ten factors influencing the corporation credit, respectively.

\end{itemize}
The rest of the paper is described below. We present
some related work in Section 2. We interpret data sets in Section 3. We introduce  models and algorithms  in section 4. We implement our  model and show the prediction results in Section 5.We discuss the  experimental results in section 6.We extract dominant factors influencing the credit rating via Factor Analysis in section 7. Finally, we draw a conclusion in section 8.

\section{Related Work}
Oricchi et al.{\color{blue}{\cite{{ref1}}}}described a proprietary SME credit rating model and found that the model's accuracy depends on the integrity of accounting information and behavioral information. Petropoulos et al.{\color{blue}{\cite{{ref2}}}}proposed an implicit Student's t-test Markov model  to calculate credit scores. Experimental results showed that the credit rating system can effectively predict corporate credit ratings. Abiyev{\color{blue}{\cite{{ref3}}}}proposed a fuzzy neural network (FNN) credit rating model to analyze corporation credit rating and evaluate its effectiveness.

Arora et al.{\color{blue}{\cite{{ref4}}}}used the random forest to classify credit indicators, and the results show that the random forest model had an excellent classification effect. Yeh et al.{\color{blue}{\cite{{ref5}}}}combined random forest and rough set theory to extract useful credit rating information. Jiang et al.{\color{blue}{\cite{{ref6}}}}proposed a support vector machine model based on Mahalanobis distance and applied it to credit risk assessment. Experimental results show that the model can be successfully used for large-scale credit risk assessment problems and had good risk assessment capabilities.

Tansel et al.{\color{blue}{\cite{{ref7}}}} used a multi-objective credit evaluation model to evaluate corporate credit and found that the multi-objective model can  accurately assess corporate risks. Addo et al.{\color{blue}{\cite{{ref8}}}}established a binary classifier model to predict the default probability of corporate loans and select the top ten important features that affected corporate credit ratings. The final experimental results showed that the model of the layered artificial neural network was more stable. Nehrebecka{\color{blue}{\cite{{ref9}}}}used logistic regression and support vector machines to assess the credit risk of non-financial companies and found that the discriminatory logistic regression model was significantly different from the support vector machine. The logistic regression model had a higher credit risk prediction rate. Angilella et al.{\color{blue}{\cite{{ref10}}}}proposed a multi-criteria credit risk model called ELECTRE-TRI to identify corporate risk categories. Hajek et al.{\color{blue}{\cite{{ref11}}}}evaluated several automatic classification methods for credit rating evaluation, using a feature selection process to improve the accuracy of classification results of credit rating indicators.

\section{Data}
This paper uses two data sets as feature sets, Financial Ratio and Balance Sheet. Data was obtained from Wharton Research Data Services(WRDS) Database for all the S\&P companies. Every sector has its own operating ways and characteristics in a company and has different effect on the credit rating of a company, Therefore, we chose specifically four typical sectors, Financial Sector, Energy Sector, Health Care Sector, Consumer Discretionary Sector. Because the type of credit rating was character variable, we needed to convert it into categorical variable with two mappings, Detailed Mapping and Coarse Mapping.The mappings were shown in Table \ref{456}, Table \ref{Coarse Mapping}. Finally, Financial Ratio and Balance Sheet combined with two mappings to form four four data sets in every sector. The results were shown in Table \ref{Dataset}.

\begin{table}[H]
\setlength\tabcolsep{3.5pt}
\centering
\caption{Detailed Mapping}
 \label{accuracy}
 \label{456}
  \setlength{\tabcolsep}{5mm}{
  \begin{tabular}{cccc}
    \toprule
   
Rating	&	Mapping	&	Rating	&	Mapping	\\
\midrule
AAA	&	1	&	BB-	&	12	\\
AA+	&	2	&	B+	&	13	\\
AA	&	3	&	B	&	14	\\
AA-	&	4	&	B-	&	15	\\
A+	&	5	&	CCC+	&	16	\\
A	&	6	&	CCC	&	17	\\
A-	&	7	&	CCC-	&	18	\\
BBB+	&	8	&	CC	&	19	\\
BBB	&	9	&	C	&	20	\\
BBB-	&	10	&	RD	&	21	\\

 \bottomrule
\end{tabular}
}
\end{table}

\begin{table}[H]
\setlength\tabcolsep{3.5pt}
\centering

     \caption{Coarse Mapping}
     \label{Coarse Mapping}

  \setlength{\tabcolsep}{5mm}{
  \begin{tabular}{ccc}
    \toprule

Rating	&	Mapping	&	Rating Description	\\
\midrule
AAA	&	1	&	Prime	\\
AA+, AA, AA-	&	2	&	High Grade	\\
A+,A,A-	&	3	&	Upper Medium Grade	\\
BBB+. BBB .BBB-	&	4	&	Lower Medium Grade	\\
BB+, BB. BB-	&	5	&	N on-Investment Grade Speculative	\\
B+, B. B-	&	6	&	Highly Speculative	\\
CCC+, CCC, CCC-	&	7	&	Substantial Risks ,	\\
CC	&	8	&	Extremely Speculative	\\
C	&	9	&	Default Imminent	\\
RD, SD, D	&	10	&	In Default	\\

 \bottomrule
\end{tabular}
}
\end{table}

\begin{table}[H]
\centering
\setlength\tabcolsep{3.5pt}
 \caption{Dataset}
     \label{Dataset}
  \setlength{\tabcolsep}{3mm}{
  \begin{tabular}{ccccc}
    \toprule

 	\quad & Data 1	&	Data 2	&	Data 3	&	Data 4	\\
 	 \midrule
Data &\tabincell{c}{	Financial\\ Ratio+\\Detailed\\ Mapping}	&\tabincell{c}{	Financial\\ Ratio+\\Coarse\\ Mapping}	&\tabincell{c}{	Balance \\Sheet+\\Detailed\\ Mapping	}&\tabincell{c}{	Balance\\ Sheet+\\Coarse \\Mapping}	\\

 \bottomrule
\end{tabular}
}

\end{table}

\section{Algorithm}
\subsection{Algorithm 1}

Principal Component Analysis (PCA) using the idea of dimension reduction (linear transformation), a set of variables of possibly related  are converted into a set of values of linearly uncorrelated variables  (principal component) under the premise of less loss of information.The first principal component has the largest possible variance (taking into account the variability of the data), and each subsequent component has the highest possible variance under the constraint of orthogonal with the previous component.The resulting vectors (each of which is a linear combination of variables and contain n variables) are unrelated orthogonal basis sets. Each principal component is a linear combination of the original variables, and the principal components are not related to each other. In general, people select principal components with cumulative proportion of variance more than 85\% because most of  information is   included. 

The top reasons for Principal component analysis are:
\begin{itemize}
    \item There are same numbers of variables and components in PCA. It can reduce the numbers of components if we just extract the components with variance cumulative proportion more than 85\% that makes the follow-up analysis easier.
    \item The objective economic phenomenon is scientifically evaluated by calculating the score of the comprehensive principal component function.
    \item It focuses on the comprehensive evaluation of the influence of information contribution.
\end{itemize}

\begin{equation}\label{eqn2}
  \begin{split}
     X_1&=a_{11}F_1+a_{12}F_2+\cdots+a_{1n}F_n \\
     X_2&=a_{21}F_1+a_{22}F_2+\cdots+a_{2n}F_n \\
     \cdots \\
     X_n&=a_{n1}F_1+a_{n2}F_2+\cdots+a_{nn}F_n   
    \end{split}
\end{equation}

 $X_n$ is the $n_{th}$ features. $F_n$ is the $n_{th}$ components. $a_{nn}=\sqrt{\lambda_n}e_{nn}$ is the loadings of components . ${\lambda_n}$  is  the variance of $n_{th}$ component(also eigenvalue). ${ \mid\mid e_n \mid\mid  =1 }$ is the  eigenvector of $n_{th}$ component.\\
 
Before performing the PCA, we must standardize data  to avoid the impact of the unit different.And in this standardized data set, the eigenvectors are  unit eigenvectors.  Therefore, we can regard ${e_{ni}}^2$ as the  proportion of variance of $i_{th}$ feature of $n_{th}$ component .We select  components with cumulative proportion of variance more than 85\%.\\ ( count: the number of components when  cumulative proportion of variance is more than 85 \%)
\begin{equation}\label{eqn2}
  \begin{split}
     X_1&=a_{11}F_1+a_{12}F_2+\cdots+a_{1count}F_{count} \\
     X_2&=a_{21}F_1+a_{22}F_2+\cdots+a_{2count}F_{count} \\
     \cdots \\
     X_n&=a_{n1}F_1+a_{n2}F_2+\cdots+a_{ncount}F_{count}   
    \end{split}
\end{equation}

The loadings of PCA represent the relationship between components and features.The higher the loading, the more the relationship between the two variables. Therefore, we consider adding the loadings of every component of each feature to get the loading scores that represents the extent to which features  affect .  Then the feature set  is  sorted in the decreasing order of components scores.In every iteration,the features from this set are  added one  in order at a time, while maintaining the order, to form the data set on which the Random Forest Classifier is trained and tested on and the accuracy of each iteration is recorded. \\

 There are  positive and negative loadings that means the loadings have positive or negative effect on components. The added scores would be counteracted if the  features had  positive and negative  loadings that  we cannot test  the importance of features.  To avoid this case, we propose two algorithms :\\
\begin{equation}\label{Algorithm 1 }
\begin{split}
 Algorithm 1 : Score_i=&\sum^{count}_{j=1}\mid a_{ij}\mid=\mid\sqrt{\lambda_1}e_{i1}\mid+\mid\sqrt{\lambda_2}e_{i2}\mid\\
 +&\cdots+\mid\sqrt{\lambda_{count}}e_{icount}\mid
  \end{split}
\end{equation}
\begin{equation}\label{Algorithm 2 }
  \begin{split}
Algorithm 2 : Score_i=\sum^{count}_{j=1} a_{ij}^2={\lambda_1}e_{i1}^2+{\lambda_2}e_{i2}^2+\cdots+{\lambda_{count}}e_{icount}^2
    \end{split}
\end{equation}

Algorithm 1 is to add  the absolute value of loading of every component of each features,the Algorithm 2  is to add  the  square of loading. They all can avoid positive and negative offsetting and represent that  features have effect on components.In order to figure out which algorithm is better,we test the accuracy of the credit rating change of Financial Data and Balance Sheet.

   \begin{figure}[H]
	\centering
		\caption{Algorithm Comparison in Detailed Mapping}
		\label{111}
	\subfigure[Financial Sector+Detailed Mapping ]{
		\begin{minipage}{0.5\linewidth}
			\centering
			\includegraphics[width=2.2in]{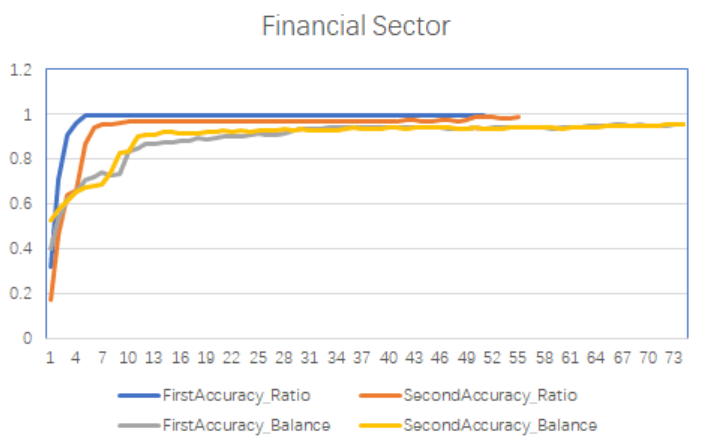}
		\end{minipage}%
	}%
	\subfigure[Energy Sector+Detailed Mapping ]{
		\begin{minipage}{0.5\linewidth}
			\centering
			\includegraphics[width=2.2in]{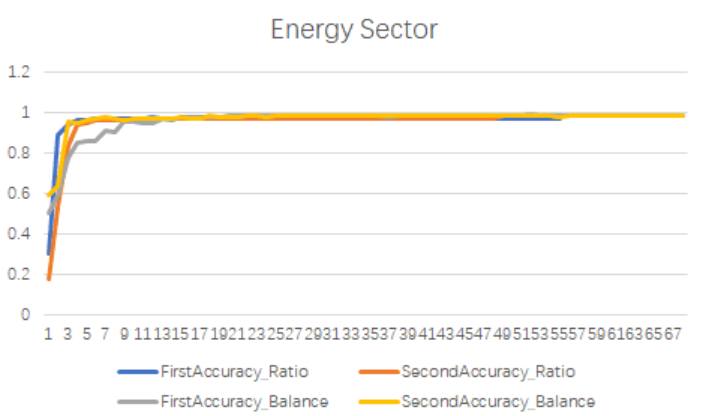}
		\end{minipage}%
	}%
\\
	\subfigure[Health Sector+Detailed Mapping ]{
		\begin{minipage}[t]{0.5\linewidth}
			\centering
			\includegraphics[width=2.2in]{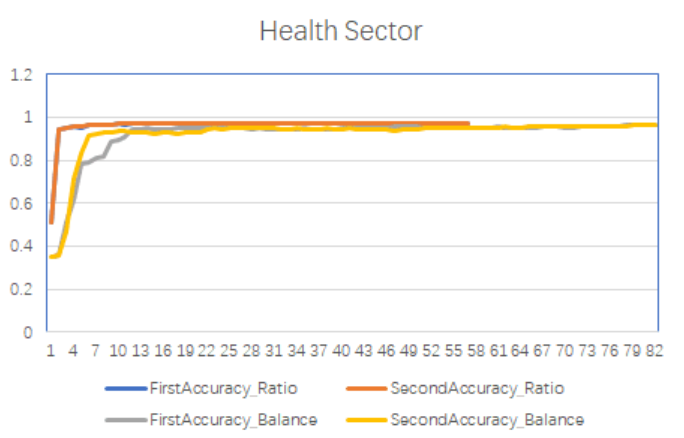}
		\end{minipage}
	}%

	\subfigure[Consumer Sector+Detailed Mapping ]{
		\begin{minipage}[t]{0.5\linewidth}
			\centering
			\includegraphics[width=2.2in]{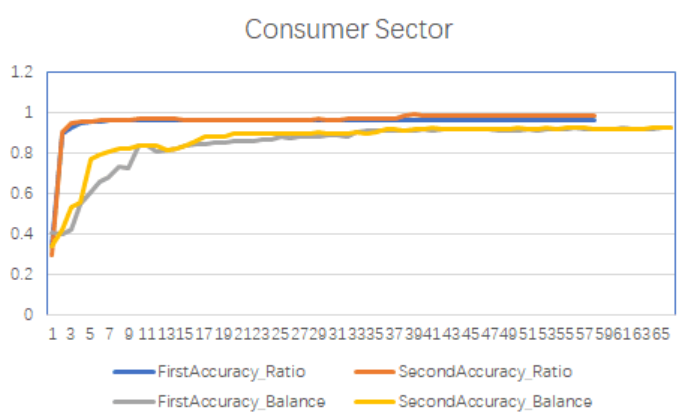}
		\end{minipage}
	}
\end{figure}
\begin{figure}[H]
\centering
		\caption{Algorithm Comparison in Coarse Mapping}
		\label{m}	
\subfigure[Financial Sector+Coarse Mapping ]{
		\begin{minipage}{0.5\linewidth}
			\centering
			\includegraphics[width=2.1in]{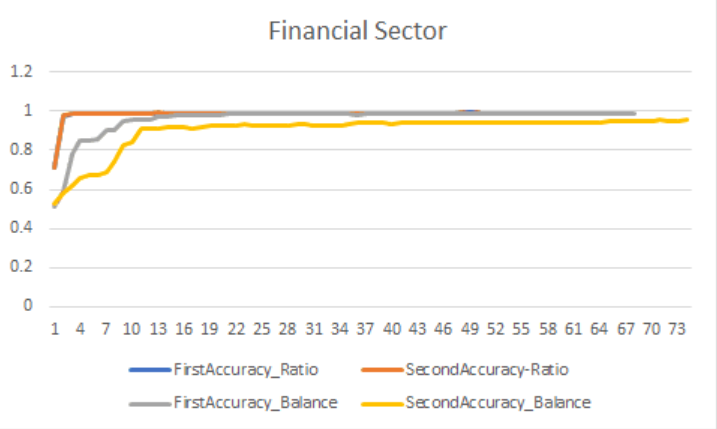}
		\end{minipage}%
	}%
	\subfigure[Energy Sector+Coarse Mapping ]{
		\begin{minipage}{0.5\linewidth}
			\centering
			\includegraphics[width=2.1in]{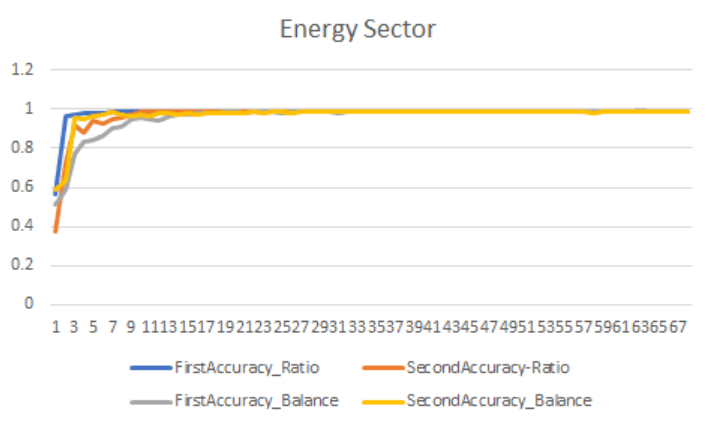}
		\end{minipage}%
	}%
\\
	\subfigure[Health Sector+Coarse Mapping]{
		\begin{minipage}[t]{0.5\linewidth}
			\centering
			\includegraphics[width=2.1in]{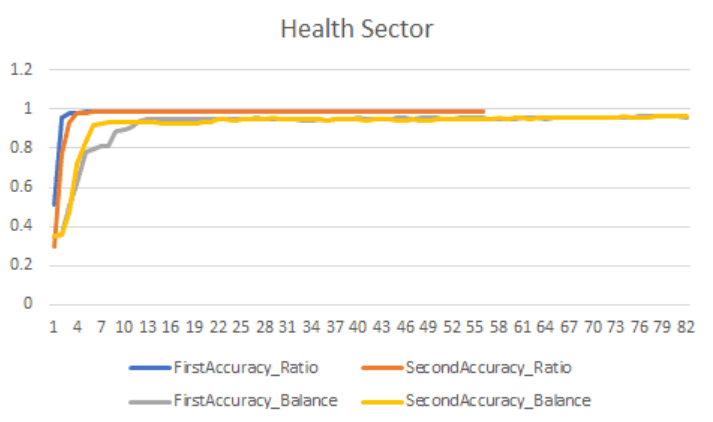}
		\end{minipage}
	}%
	\subfigure[Consumer Sector+Coarse Mapping]{
		\begin{minipage}[t]{0.5\linewidth}
			\centering
			\includegraphics[width=2.1in]{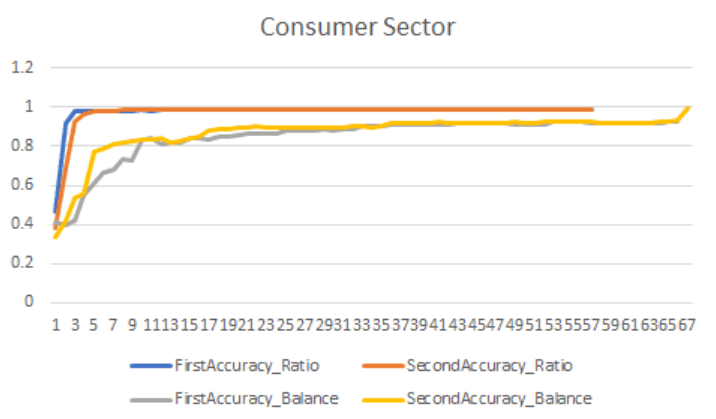}
		\end{minipage}
	}
\end{figure}

From Figure \ref{111}, \ref{m}, the blue and gray line show the accuracy of  algorithm 1 in Financial Ratio and Balance Sheet respectively;orange  and yellow line display the accuracy of  algorithm 2 in Financial Ratio and Balance Sheet respectively.In Financial Ratio, there is slight difference between these two algorithms in four sectors. In Balance Sheet,yellow line (algorithm 2)  is always above the gray line(algorithm 1) that means the accuracy of algorithm 2 is more accurate than that of algorithm 1 in four sectors.\\

However, the numbers of selected features for four data sets in every sector are different from each other  because we select the features with Chi-square p-value less than 0.05  as final feature sets for four data sets in every sector. So we try to test the accuracy with the  same numbers. From Figure \ref{111}, \ref{m} ,most accuracy curves tent to steady when the number is  20 or more. We calculate  the average accuracy of the top 20 features by utilizing Random Forest that trained and tested the accuracy of feature sets for 100 times. The results are shown in Table \ref{222}, Table \ref{333}, Table \ref{444}, Table \ref{555}.
\begin{table}[H]
\centering\scriptsize
\caption{Accuracy for Financial Sector}
\label{222}
\setlength\tabcolsep{3.5pt}
\begin{tabular}{ccccc}
\hline\hline\noalign{\smallskip}
\quad & \tabincell{c}{Financial Raio\\Detailed Mapping} &\tabincell{c}{Balance Sheet\\Detailed Mapping}&\tabincell{c}{ Financial Ratio\\Coarse Mapping }&\tabincell{c}{ Balance Sheet\\Coarse Mapping } \\
\noalign{\smallskip}\hline\noalign{\smallskip}
		\tabincell{c}{Avg.Test Accuracy\\ Accuracy with\\ Algorithm 1 using \\ The Top 20 Features} & 0.97169 & 0.87244& 0.98999 & 0.8964\\
		\specialrule{0em}{1.5pt}{1.5pt}
			\tabincell{c}{Avg.Test Accuracy\\Accuracy with \\ Algorithm 2 using\\ The Top 20 Features} &0.97577 &0.92332 & 0.98989 &0.9274 \\
      		\noalign{\smallskip}\hline\hline	
	\end{tabular}
	\end{table}
\begin{table}[H]
\centering\scriptsize
\caption{Accuracy for Energy Sector}
\label{333}
\setlength\tabcolsep{3.5pt}
\begin{tabular}{ccccc}	
		\hline\hline\noalign{\smallskip}
		\quad & \tabincell{c}{Financial Raio\\Detailed Mapping} &\tabincell{c}{Balance Sheet\\Detailed Mapping}&\tabincell{c}{ Financial Ratio\\Coarse Mapping }&\tabincell{c}{ Balance Sheet\\Coarse Mapping } \\
		\noalign{\smallskip}\hline\noalign{\smallskip}
		\tabincell{c}{Avg.Test Accuracy\\ Accuracy with \\ Algorithm 1 using \\ The Top 20 Features} & 0.97271 &0.90403 & 0.98915 & 0.98205\\
		\specialrule{0em}{1.5pt}{1.5pt}
			\tabincell{c}{Avg.Test Accuracy\\Accuracy with \\ Algorithm 2 using\\ The Top 20 Features} &0.97035 &0.91974 & 0.98947 & 0.98064\\		
		\noalign{\smallskip}\hline\hline	
	\end{tabular}
	\end{table}
\begin{table}[H]
\centering\scriptsize
\caption{Accuracy for Health Sector}
\label{444}
\setlength\tabcolsep{3.5pt}
\begin{tabular}{ccccc}		
		\hline\hline\noalign{\smallskip}
		\quad & \tabincell{c}{Financial Raio\\Detailed Mapping} &\tabincell{c}{Balance Sheet\\Detailed Mapping}&\tabincell{c}{ Financial Ratio\\Coarse Mapping }&\tabincell{c}{ Balance Sheet\\Coarse Mapping } \\
		\noalign{\smallskip}\hline\noalign{\smallskip}
		\tabincell{c}{Avg.Test Accuracy\\ Accuracy with \\ Algorithm 1 using\\ The Top 20 Features} & 0.97083 &0.86841 & 0.99001 & 0.95187\\
		\specialrule{0em}{1.5pt}{1.5pt}
			\tabincell{c}{Avg.Test Accuracy\\Accuracy with \\ Algorithm 2 using\\ The Top 20 Features} &0.97288 &0.93562 & 0.99035 & 0.93103\\		
		\noalign{\smallskip}\hline\hline	
	\end{tabular}
	\end{table}

\begin{table}[H]
\centering\scriptsize
\caption{Accuracy for Consumer Sector}
\label{555}
\setlength\tabcolsep{3.5pt}
\begin{tabular}{ccccc}	
		\hline\hline\noalign{\smallskip}
		\quad & \tabincell{c}{Financial Raio\\Detailed Mapping} &\tabincell{c}{Balance Sheet\\Detailed Mapping}&\tabincell{c}{ Financial Ratio\\Coarse Mapping }&\tabincell{c}{ Balance Sheet\\Coarse Mapping } \\
		\noalign{\smallskip}\hline\noalign{\smallskip}
		\tabincell{c}{Avg.Test Accuracy\\ Accuracy with \\ Algorithm 1 using\\ The Top 20 Features} & 0.96281 & 0.80097& 0.98614 & 0.86128\\
			\specialrule{0em}{1.5pt}{1.5pt}
			\tabincell{c}{Avg.Test Accuracy\\Accuracy with \\ Algorithm 2 using\\ The Top 20 Features} & 0.96734&0.93562& 0.99035 &0.93103 \\	
		\noalign{\smallskip}\hline\hline	
	\end{tabular}
	\end{table}

\begin{figure}[H]
	\centering
     \caption{Flowchart for Algorithm 1}
     \label{Flowchart for Algorithm 1}
		\includegraphics[width=5in]{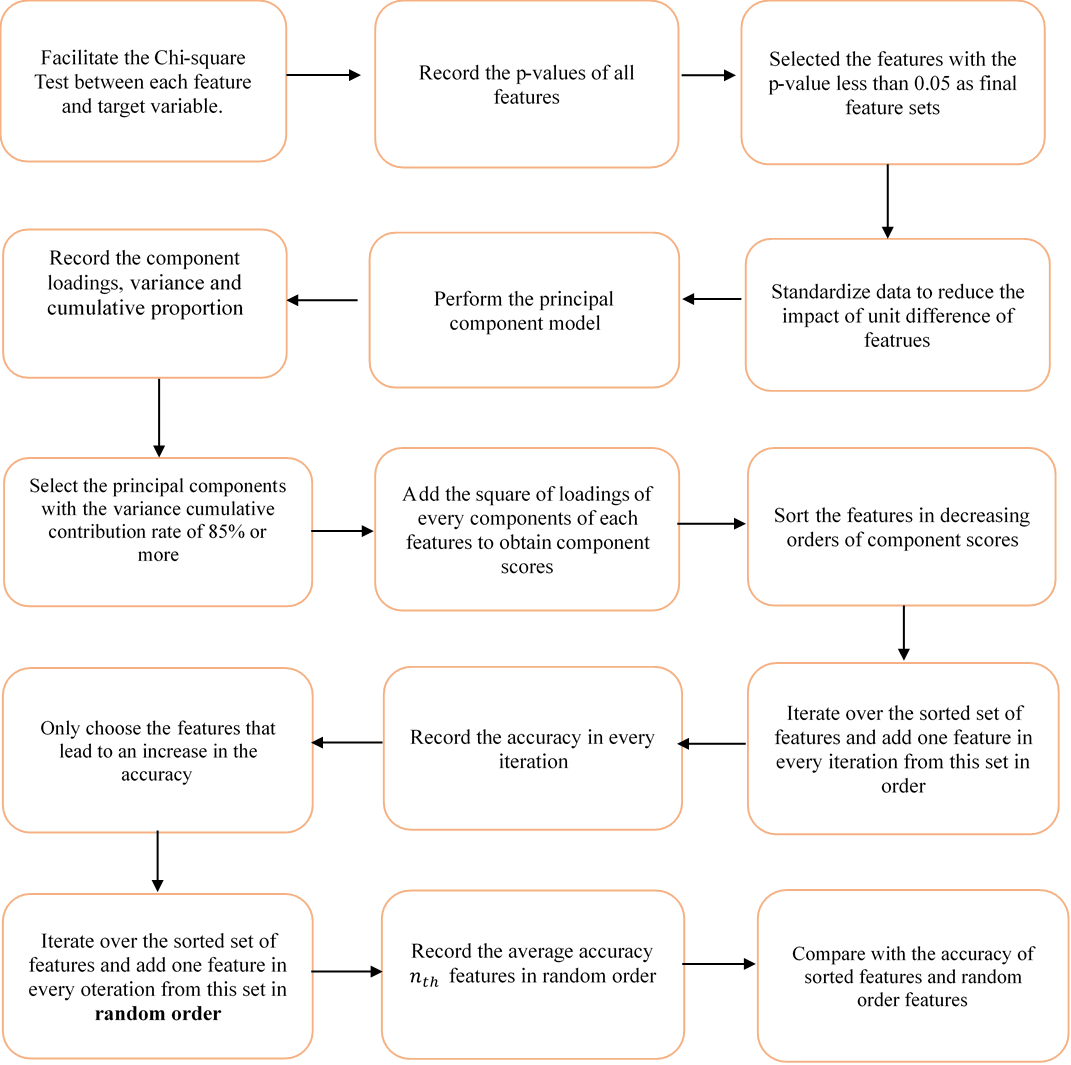}	
	\end{figure}

From Table \ref{222}, Table \ref{333}, Table \ref{444}, Table \ref{555}, most accuracy of Algorithm 2 is higher than  that of algorithm 1 in four sectors. Therefore, algorithm 2 in this part is designed for the final Algorithm 1.After obtaining  the final  algorithm 1, a summary of  this algorithm is shown in Figure \ref{Flowchart for Algorithm 1}. 
	
	\newpage
	
\subsection{Algorithm 2} \label{sec:sampling}
 Factor analysis is to identify the underlying relationships between measured variables. And Factor analysis is related to principal component analysis (PCA), but the two are not identical.Factor Analysis is based on the idea of dimension reduction and the study of the dependence within the correlation matrix of the original variables, some variables with complex relationships are expressed as a linear combination of a small number of common factors and special factors that only work on one variable.In this algorithm, we try to summarize the similar features together to extract  common factors that can interpret features  better.

The top reasons for Factor analysis are:
\begin{itemize}
    \item  Factor analysis make the factor better explained by exploiting factor rotation. The factor analysis is more dominant in the interpretation of the principal component.
    \item Factor analysis is not to delete  original variables ,but the information of the original variable is recombined to find a common factor affecting the target variable and  simplify data.
\end{itemize}

Before performing Factor Analysis, feature set must be tested by KMO and Bartlett's Test to examine whether the  data is suitable for factor analysis.\\

The Kaiser-Meyer-Olkin (KMO) test measures the applicability of our data to factor analysis. The test measures sampling adequacy for each variable in the model and for the complete model.KMO values less than 0.6 indicate the sampling is not adequate that factor analysis  should not be taken.\\

Bartlett's Test of Sphericity is based on the correlation coefficient matrix. Its null hypothesis correlation coefficient matrix is a unit array, that is, all the elements of the diagonal of the correlation coefficient matrix are 1, and all the non-diagonal elements are zero. The statistics of Bartlett's Test of Sphericity is derived from the determinant of the correlation coefficient matrix.If the value is large, and the corresponding concomitant probability value is less than the specified significant level, the null hypothesis is rejected, indicating that the correlation coefficient matrix is not a unit array, there is a correlation between the original variables. Small p-value (less than 0.05) of the significance level indicate that a factor analysis may be useful with  data.

 \begin{figure}[H]
	\centering
	\setlength{\abovecaptionskip}{0.05cm}
     \setlength{\belowcaptionskip}{-0.05cm} 
	
	\subfigure[Financial Ratio]{
		\begin{minipage}{0.5\linewidth}
			\centering
			\includegraphics[width=2.2in]{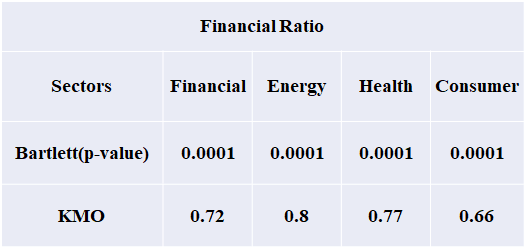}
		\end{minipage}%
	}%
	\subfigure[Balance Sheet]{
		\begin{minipage}{0.5\linewidth}
			\centering
			\includegraphics[width=2.2in]{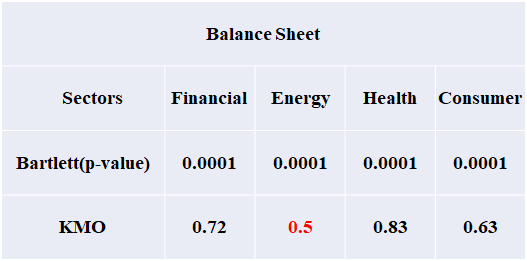}
		\end{minipage}%
	}%
		\caption{KMO and Bartlett's Test}
		\label{KMO}
\end{figure}

From Figure \ref{KMO}, almost  all the p-values of  Bartlett's Test are less than 0.05 and KMO vaules are more than 0.6 in four sectors in Financial Ratios and Balance Sheet except for Energy Sector in Balance Sheet.

\begin{figure}[H]
	\centering
     \caption{Flowchart for Algorithm 2}
     \label{Flowchart for Algorithm 2}
		\includegraphics[width=5in]{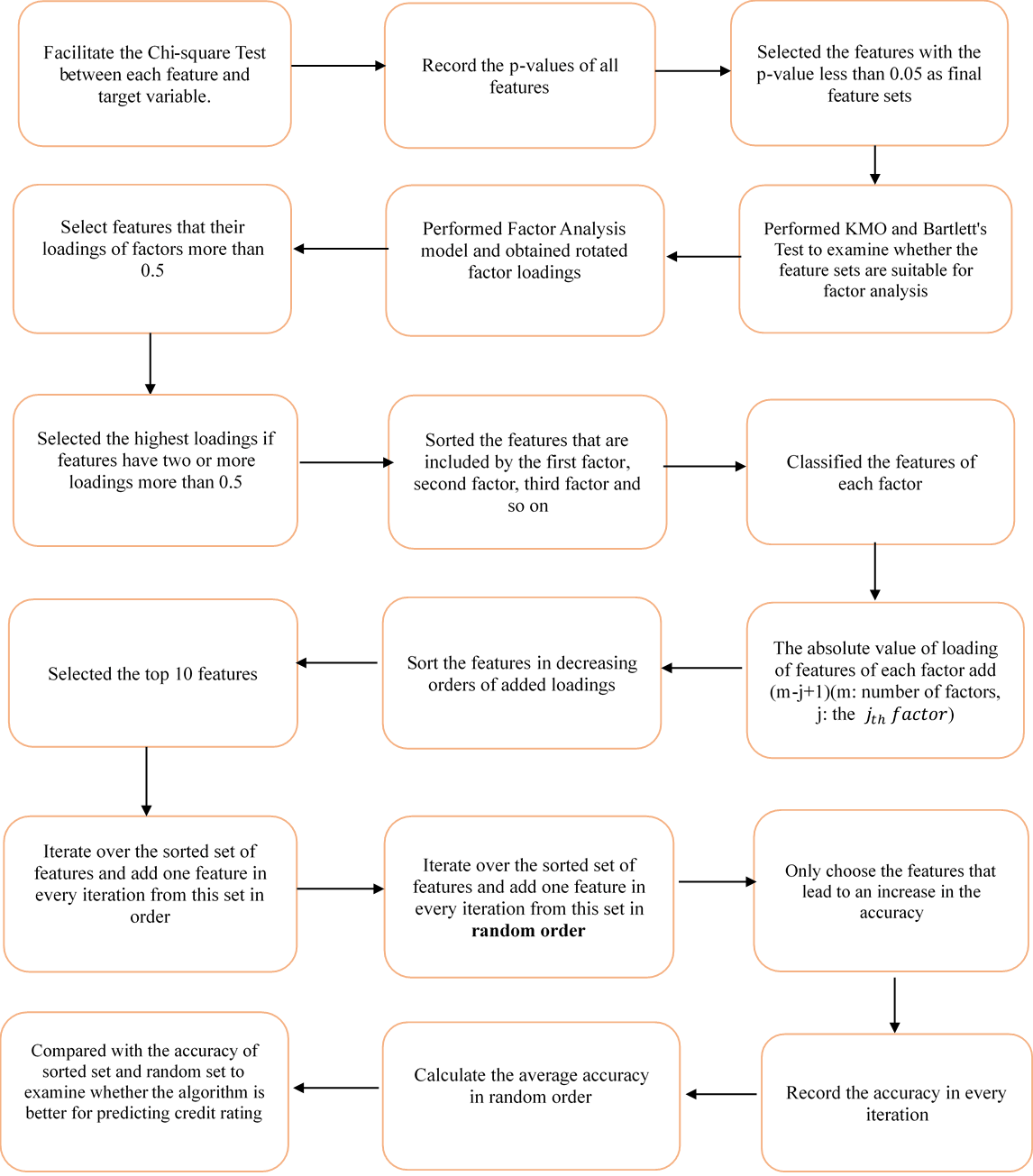}	
	\end{figure}

After performing the Factor Analysis to get the rotated loadings, the  features with loadings  more than 0.5 are selected and the features are deleted if  their loadings with less than 0.5.  The  feature represents the factor that its loading is maximum if   they have two or more loadings more than 0.5. Because the factors are unrelated to  each other, each feature can be  only included by one factor,there are no same features between  factors. The loading of features of factor means that have psitive or negative effect on factor and  there is first factor,second factor and so on with their effects gradually weakening. According to the principle of priority of the first factor ,  the features of each factor are sorted according to the  absolute value  of loadings within first factor and then second factor and so on. For example,  if first factor includes three features, their order are 1,2,3; then second factors has four features,their order are 4,5,6,7 and so on. To obtain this sorted set, the algorithm 2 is proposed. If there are m factors, feature A is included by the $j_{th}$ factor, the absolute value of loading of A plus (m-j+1) to get factor scores. It means that absolute value  of loadings of the features of  first factor plus highest score m, that  of the second factor plus  second highest score (m-1),that  of the $j_{th}$  factor plus   score  (m-j+1) and so on. Then the feature set  is  sorted in the decreasing order of added scores. In every iteration,the features from this set are  added one  in order at a time, while maintaining the order, to form the data set on which the Random Forest Classifier is trained and tested on and the accuracy of each iteration is recorded. A summary of  this algorithm is shown in Figure \ref{Flowchart for Algorithm 2}. 
 
\setcounter{table}{0}

\setcounter{figure}{0}
\newpage
\section{Prediction Results}

In Algorithm 1,  the components with cumulative variance proportion more than 85\%  were selected as final principal components.The number of component was shown in Table \ref{The Number of Component for Algorithm 1 }.  In  Algorithm 2, the factors  with cumulative variance proportion more than 85\%  were selected as final factors. The number of factor was shown in Table  \ref{The Number of Factors for Algorithm 2}.\\

In this section, the results for both the algorithms will be summarized according to  sectors. We compared  the accuracy of the  top 20 features selected by Algorithm 1 and the accuracy of the features selected by Algorithm 2. These two feature sets were trained and tested  for 100 times by utilizing Random Forest Classifier and the average accuracy was  recorded. The results were shown in Table \ref{Accuracy Results for Financial Sector}, Table \ref{Accuracy Results for Energy Sector}, Table \ref{Accuracy  Results for Health  Sector}, Table \ref{Accuracy Results for Consumer  Sector}.\\

\begin{table}[H]
\centering
\caption{The Number of Component for Algorithm 1 }
\label{The Number of Component for Algorithm 1 }
\setlength\tabcolsep{3.5pt} 
\scriptsize
\begin{tabular}{ccccc  }
		
		\hline\hline\noalign{\smallskip}
		\quad & \tabincell{c}{Financial Raio\\Detailed Mapping} &\tabincell{c}{Financial Ratio\\Coarse Mapping}&\tabincell{c}{ Balance Sheet\\Detailed Mapping }&\tabincell{c}{ Balance Sheet\\Coarse Mapping } \\
		\noalign{\smallskip}\hline\noalign{\smallskip}
		\tabincell{c}{Financial Sector} & 22 &22 & 33& 32\\
			\specialrule{0em}{1.5pt}{1.5pt}
			\tabincell{c}{Energy Sector} &18& 18&22 &21  \\
			\specialrule{0em}{1.5pt}{1.5pt}
      \tabincell{c}{Health Sector} &22 &22 &29  &32 \\
      	\specialrule{0em}{1.5pt}{1.5pt}
      \tabincell{c}{  Consumer Sector }  &28 &29 &28  &28 \\		
		\noalign{\smallskip}\hline\hline	
	\end{tabular}
	\end{table}

\begin{table}[H]
\centering
\setlength\tabcolsep{3.5pt} 
\caption{The Number of Factors for Algorithm 2 }
\label{The Number of Factors for Algorithm 2}
\scriptsize
\begin{tabular}{ccccc  }
		
		\hline\hline\noalign{\smallskip}
		\quad & \tabincell{c}{Financial Raio\\Detailed Mapping} &\tabincell{c}{Financial Ratio\\Coarse Mapping}&\tabincell{c}{ Balance Sheet\\Detailed Mapping }&\tabincell{c}{ Balance Sheet\\Coarse Mapping } \\
		\noalign{\smallskip}\hline\noalign{\smallskip}
		\tabincell{c}{Financial Sector} & 10 &10 & 15& 14\\
			\specialrule{0em}{1.5pt}{1.5pt}
			\tabincell{c}{Energy Sector} &8& 8&NA &NA \\
			\specialrule{0em}{1.5pt}{1.5pt}
      \tabincell{c}{Health Sector} &11 &10 &11  &13 \\
      	\specialrule{0em}{1.5pt}{1.5pt}
      \tabincell{c}{  Consumer Sector }  &11 &11 &13  &12 \\
		
		\noalign{\smallskip}\hline\hline	
	\end{tabular}
	\end{table}

\begin{table}[H]
\centering
\caption{Accuracy Results for Financial Sector}
\label{Accuracy Results for Financial Sector}
\scriptsize
\setlength\tabcolsep{3pt} 
\begin{tabular}{ccccc  }
		
		\hline\hline\noalign{\smallskip}
		\quad & \tabincell{c}{Financial Raio\\Detailed Mapping} &\tabincell{c}{Balance Sheet\\Detailed Mapping}&\tabincell{c}{ Financial Ratio\\Coarse Mapping }&\tabincell{c}{ Balance Sheet\\Coarse Mapping } \\
		\noalign{\smallskip}\hline\noalign{\smallskip}
		\tabincell{c}{Avg.Test Accuracy\\ Accuracy with \\ Algorithm 1 using\\ The Top 20 Features} & 0.97577 &0.92332 & 0.98989 & 0.9274\\
			\specialrule{0em}{1.5pt}{1.5pt}
			\tabincell{c}{Avg.Test Accuracy\\ Accuracy with\\ Algorithm  2} &0.97465 &0.92331 & 0.98983 &0.95227 \\
			\specialrule{0em}{1.5pt}{1.5pt}
      	\specialrule{0em}{1.5pt}{1.5pt}
      \tabincell{c}{  Avg.No feature\\ Selection  } &0.97712 & 0.92766 & 0.99203&0.96721\\
		
		\noalign{\smallskip}\hline\hline
	
	\end{tabular}
	\end{table}

\begin{table}[H]
\centering
\setlength\tabcolsep{3pt} 
\caption{Accuracy Results for Energy Sector}
\label{Accuracy Results for Energy Sector}
\scriptsize
\begin{tabular}{ccccc}
		\hline\hline\noalign{\smallskip}
		\quad & \tabincell{c}{Financial Raio\\Detailed Mapping} &\tabincell{c}{Balance Sheet\\Detailed Mapping}&\tabincell{c}{ Financial Ratio\\Coarse Mapping }&\tabincell{c}{ Balance Sheet\\Coarse Mapping } \\
		\noalign{\smallskip}\hline\noalign{\smallskip}
		\tabincell{c}{Avg.Test Accuracy\\ Accuracy with \\ Algorithm 1 using\\ The Top 20 Features} & 0.97035 & 0.91974&0.98947  & 0.98064\\
		\specialrule{0em}{1.5pt}{1.5pt}
			\tabincell{c}{Avg.Test Accuracy\\ Accuracy with\\ Algorithm  2} &0.96859 &NA &0.98831  & NA\\
			\specialrule{0em}{1.5pt}{1.5pt}
     \specialrule{0em}{1.5pt}{1.5pt}
      \tabincell{c}{  Avg.No feature\\ Selection  } & 0.97253& 0.93115 & 0.99027&0.98647\\		
		\noalign{\smallskip}\hline\hline	
	\end{tabular}
	\end{table}

\begin{table}[H]
\centering
\caption{Accuracy  Results for Health Sector}
\label{Accuracy  Results for Health  Sector}
\setlength\tabcolsep{3pt} 
\scriptsize
\begin{tabular}{ccccc}	
		\hline\hline\noalign{\smallskip}
		\quad & \tabincell{c}{Financial Raio\\Detailed Mapping} &\tabincell{c}{Balance Sheet\\Detailed Mapping}&\tabincell{c}{ Financial Ratio\\Coarse Mapping }&\tabincell{c}{ Balance Sheet\\Coarse Mapping } \\
		\noalign{\smallskip}\hline\noalign{\smallskip}
		\tabincell{c}{Avg.Test Accuracy\\ Accuracy with \\ Algorithm 1 using\\ The Top 20 Features} & 0.97288 &0.93562 & 0.99035 & 0.93103\\
		\specialrule{0em}{1.5pt}{1.5pt}
			\tabincell{c}{Avg.Test Accuracy\\ Accuracy with\\ Algorithm  2} &0.97363 & 0.93091& 0.99081 & 0.96093\\
			\specialrule{0em}{1.5pt}{1.5pt}
     \specialrule{0em}{1.5pt}{1.5pt}
      \tabincell{c}{  Avg.No feature\\ Selection  } &0.97402 &0.94367  &0.99112 &0.96389\\	
		\noalign{\smallskip}\hline\hline	
	\end{tabular}
	\end{table}
\begin{table}[H]
\centering
\caption{Accuracy  Results for Consumer Sector}
\label{Accuracy Results for Consumer  Sector}
\setlength\tabcolsep{3pt} 
\scriptsize
\begin{tabular}{ccccc}	
		\hline\hline\noalign{\smallskip}
		\quad & \tabincell{c}{Financial Raio\\Detailed Mapping} &\tabincell{c}{Balance Sheet\\Detailed Mapping}&\tabincell{c}{ Financial Ratio\\Coarse Mapping }&\tabincell{c}{ Balance Sheet\\Coarse Mapping } \\
		\noalign{\smallskip}\hline\noalign{\smallskip}
		\tabincell{c}{Avg.Test Accuracy\\ Accuracy with \\ Algorithm 1 using\\ The Top 20 Features} &0.96734  &0.93562& 0.99035 &0.93103 \\
			\specialrule{0em}{1.5pt}{1.5pt}
			\tabincell{c}{Avg.Test Accuracy\\ Accuracy with\\ Algorithm  2} &0.96784 & 0.93091& 0.99081 & 0.96093\\
			\specialrule{0em}{1.5pt}{1.5pt}
      	\specialrule{0em}{1.5pt}{1.5pt}
      \tabincell{c}{  Avg.No feature\\ Selection  } & 0.96836& 0.94367 & 0.99112&0.96389\\		
		\noalign{\smallskip}\hline\hline	
	\end{tabular}
	\end{table}
 \setcounter{table}{0}

\setcounter{figure}{0}
 
\section{Discussion }

We can see from Table \ref{Accuracy Results for Financial Sector}, \ref{Accuracy Results for Energy Sector}, \ref{Accuracy  Results for Health  Sector}, \ref{Accuracy Results for Consumer  Sector} , they showed the similar results in term of the accuracy of Algorithms. However,the number of selected features  by them are different. For all the original feature sets,the features with the p-value less than 0.05 were selected by exploiting Chi-square Test. Algorithm1 was to add the loading square of every component  for each features  as component scores of each features and sort the features according to the compnent scores,it did not need to delete features.Therefore the number of feature sets of Algorithm 1 was equal to that of Chi-Suqare Test. Algorithm 2 was to delete features that had no loadings with more than 0.5 and sort the deleted feature sets. Therefore the number of feature sets of Algorithm 2 was not equal to that of Chi-Suqare Test.The results are shown in Table \ref{The Number of original Features }, \ref{The Number of  Features selected by Chi-Square Test }, \ref{The Number of  Features for  Alogorithm 1  } , \ref{The Number of  Features selected for  Alogorithm 2 }.

\begin{table}[H]
\centering
\caption{The Number of original Features }
\label{The Number of original Features }
\scriptsize
\setlength\tabcolsep{3pt} 
\begin{tabular}{ccccc}		
		\hline\hline\noalign{\smallskip}
		\quad & \tabincell{c}{Financial Raio\\Detailed Mapping} &\tabincell{c}{Financial Ratio\\Coarse Mapping}&\tabincell{c}{ Balance Sheet\\Detailed Mapping }&\tabincell{c}{ Balance Sheet\\Coarse Mapping } \\
		\noalign{\smallskip}\hline\noalign{\smallskip}
		\tabincell{c}{Financial Sector} & 52 &52 & 101& 101\\
			\specialrule{0em}{1.5pt}{1.5pt}
			\tabincell{c}{Energy Sector} &57& 57&111 &111   \\
			\specialrule{0em}{1.5pt}{1.5pt}
      \tabincell{c}{Health Sector} &59 &59 &116  &116 \\
      	\specialrule{0em}{1.5pt}{1.5pt}
      \tabincell{c}{  Consumer Sector }  &58 &58 &103  &103 \\		
		\noalign{\smallskip}\hline\hline	
	\end{tabular}

	\end{table}
\begin{table}[H]
\centering
\caption{The Number of  Features selected by Chi-Square Test}
\setlength\tabcolsep{3pt} 
\label{The Number of  Features selected by Chi-Square Test }
\scriptsize
\begin{tabular}{ccccc  }		
		\hline\hline\noalign{\smallskip}
		\quad & \tabincell{c}{Financial Raio\\Detailed Mapping} &\tabincell{c}{Financial Ratio\\Coarse Mapping}&\tabincell{c}{ Balance Sheet\\Detailed Mapping }&\tabincell{c}{ Balance Sheet\\Coarse Mapping } \\
		\noalign{\smallskip}\hline\noalign{\smallskip}
		\tabincell{c}{Financial Sector} & 51 &50& 76& 74\\
			\specialrule{0em}{1.5pt}{1.5pt}
			\tabincell{c}{Energy Sector} &54& 55&71 &68  \\
			\specialrule{0em}{1.5pt}{1.5pt}
      \tabincell{c}{Health Sector} &57 &56 &75 &82 \\
      	\specialrule{0em}{1.5pt}{1.5pt}
      \tabincell{c}{  Consumer Sector }  &58 &57 &73  &66 \\		
		\noalign{\smallskip}\hline\hline	
	\end{tabular}
	\end{table}

\begin{table}[H]
\centering
\caption{The Number of  Features for  Alogorithm 1 }
\label{The Number of  Features for  Alogorithm 1 }
\setlength\tabcolsep{3pt} 
\scriptsize
\begin{tabular}{ccccc}		
		\hline\hline\noalign{\smallskip}
		\quad & \tabincell{c}{Financial Raio\\Detailed Mapping} &\tabincell{c}{Financial Ratio\\Coarse Mapping}&\tabincell{c}{ Balance Sheet\\Detailed Mapping }&\tabincell{c}{ Balance Sheet\\Coarse Mapping } \\
		\noalign{\smallskip}\hline\noalign{\smallskip}
		\tabincell{c}{Financial Sector} & 51 &50& 76& 74\\
			\specialrule{0em}{1.5pt}{1.5pt}
			\tabincell{c}{Energy Sector} &54& 55&71 &68  \\
			\specialrule{0em}{1.5pt}{1.5pt}
      \tabincell{c}{Health Sector} &57 &56 &75 &82 \\
      	\specialrule{0em}{1.5pt}{1.5pt}
      \tabincell{c}{  Consumer Sector }  &58 &57 &73  &66 \\		
		\noalign{\smallskip}\hline\hline	
	\end{tabular}
	\end{table}

\begin{table}[H]
\centering
\caption{The Number of  Features selected for  Alogorithm 2 }
\label{The Number of  Features selected for  Alogorithm 2 }
\setlength\tabcolsep{3pt} 
\scriptsize
\begin{tabular}{ccccc}		
		\hline\hline\noalign{\smallskip}
		\quad & \tabincell{c}{Financial Raio\\Detailed Mapping} &\tabincell{c}{Financial Ratio\\Coarse Mapping}&\tabincell{c}{ Balance Sheet\\Detailed Mapping }&\tabincell{c}{ Balance Sheet\\Coarse Mapping } \\
		\noalign{\smallskip}\hline\noalign{\smallskip}
		\tabincell{c}{Financial Sector} & 31 &31& 52& 50\\
			\specialrule{0em}{1.5pt}{1.5pt}
			\tabincell{c}{Energy Sector} &37& 38&NA &NA \\
			\specialrule{0em}{1.5pt}{1.5pt}
      \tabincell{c}{Health Sector} &36 &35 &52 &57 \\
      	\specialrule{0em}{1.5pt}{1.5pt}
      \tabincell{c}{  Consumer Sector }  &29 &30 &48  &43 \\		
		\noalign{\smallskip}\hline\hline	
	\end{tabular}
	\end{table}

\begin{figure}[H]
	\centering
	
	\subfigure[Financial Sector]{
		\begin{minipage}{0.5\linewidth}
			\centering
			\includegraphics[width=2.2in]{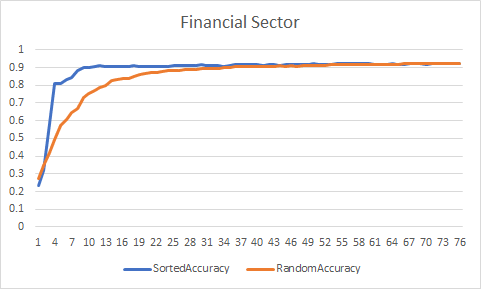}
		\end{minipage}%
	}%
	\subfigure[Energy Sector]{
		\begin{minipage}{0.5\linewidth}
			\centering
			\includegraphics[width=2.2in]{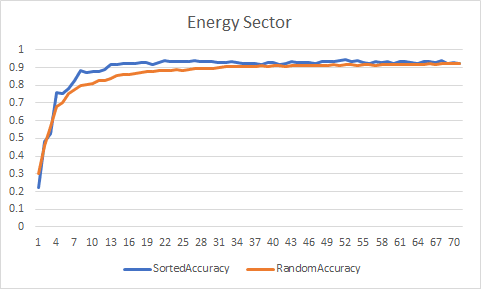}
		\end{minipage}%
	}%
\\
	\subfigure[Health Sector]{
		\begin{minipage}[t]{0.5\linewidth}
			\centering
			\includegraphics[width=2.1in]{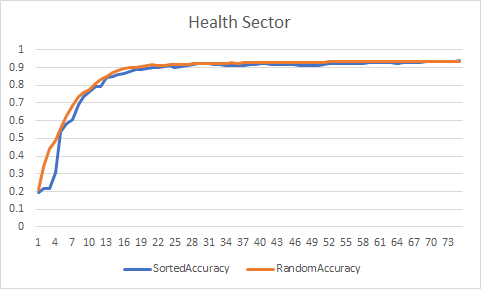}
		\end{minipage}
	}%
	\subfigure[Consumer Sector]{
		\begin{minipage}[t]{0.5\linewidth}
			\centering
			\includegraphics[width=2.1in]{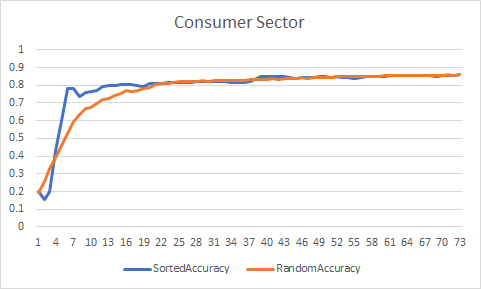}
		\end{minipage}
	}\caption{Feature Number in Balance Sheet Detailed Mapping for Algorithm 1}
		\label{Feature Number in Balance Sheet Detailed Mapping for Algorithm 1}

\end{figure}

\begin{figure}[H]
	\centering
	
	\subfigure[Financial Sector]{
		\begin{minipage}{0.5\linewidth}
			\centering
			\includegraphics[width=2.2in]{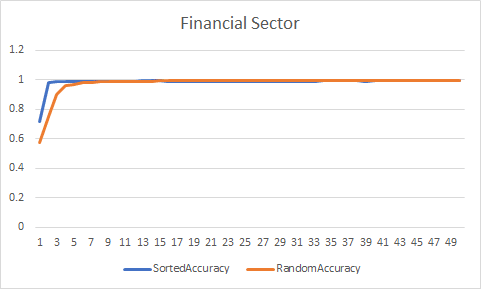}
		\end{minipage}%
	}%
	\subfigure[Energy Sector]{
		\begin{minipage}{0.5\linewidth}
			\centering
			\includegraphics[width=2.2in]{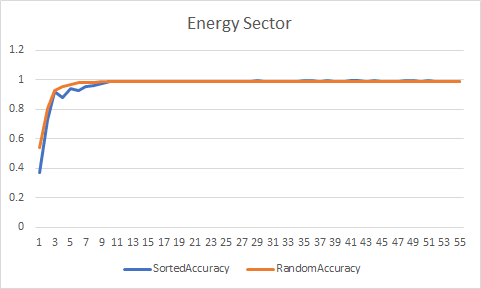}
		\end{minipage}%
	}%
\\
	\subfigure[Health Sector]{
		\begin{minipage}[t]{0.5\linewidth}
			\centering
			\includegraphics[width=2.2in]{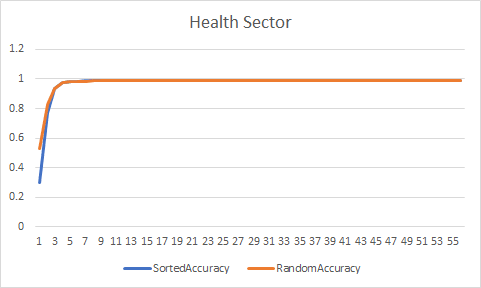}
		\end{minipage}
	}%
	\subfigure[Consumer Sector]{
		\begin{minipage}[t]{0.5\linewidth}
			\centering
			\includegraphics[width=2.2in]{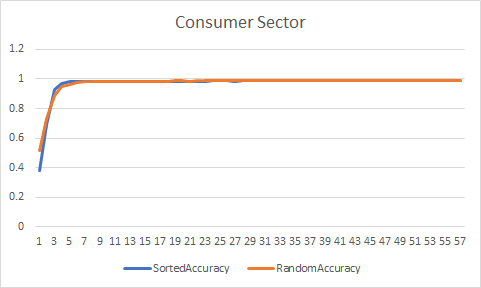}
		\end{minipage}
	}
		\caption{Feature Number in Financial Ratio Coarse Mapping for Algorithm 1 }
		\label{Feature Number in Financial Ratio+Coarse Mapping for Algorithm 1}

\end{figure}

\begin{figure}[H]
	\centering
		
	\subfigure[Financial Sector]{
		\begin{minipage}{0.5\linewidth}
			\centering
			\includegraphics[width=2.2in]{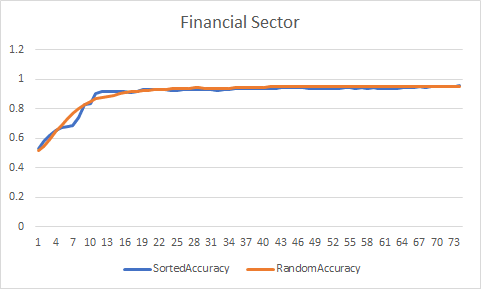}
		\end{minipage}%
	}%
	\subfigure[Health Sector]{
		\begin{minipage}{0.5\linewidth}
			\centering
			\includegraphics[width=2.2in]{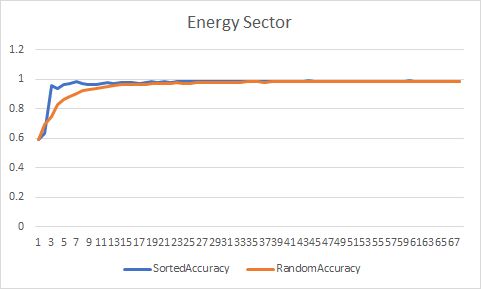}
		\end{minipage}%
	}%
\\
\subfigure[Financial Sector]{
		\begin{minipage}{0.5\linewidth}
			\centering
			\includegraphics[width=2.2in]{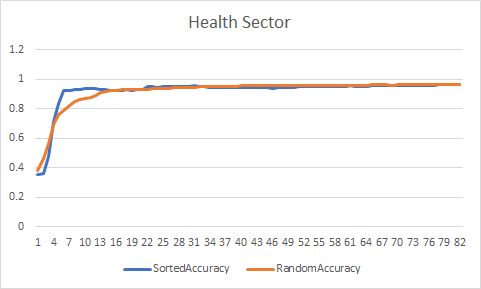}
		\end{minipage}%
	}%
	\subfigure[Health Sector]{
		\begin{minipage}{0.5\linewidth}
			\centering
			\includegraphics[width=2.2in]{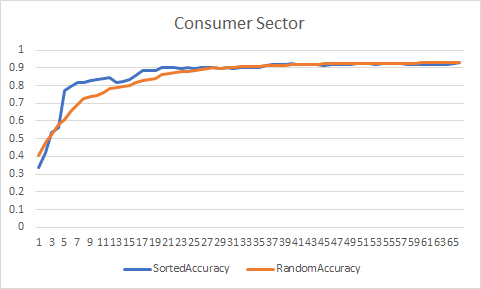}
		\end{minipage}%
	}%
	\caption{Feature Number in Balance Sheet  Coarse Mapping for Algorithm 1}
			\label{Accuracy in Balance Sheet +Coarse Mapping for Algorithm 1}

\end{figure}

\begin{figure}[H]
	\centering
		
	\subfigure[Financial Sector]{
		\begin{minipage}{0.5\linewidth}
			\centering
			\includegraphics[width=2.2in]{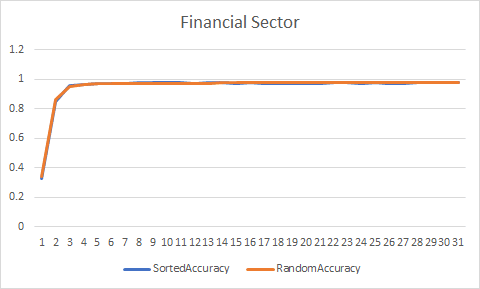}
		\end{minipage}%
	}%
	\subfigure[Energy Sector]{
		\begin{minipage}{0.5\linewidth}
			\centering
			\includegraphics[width=2.2in]{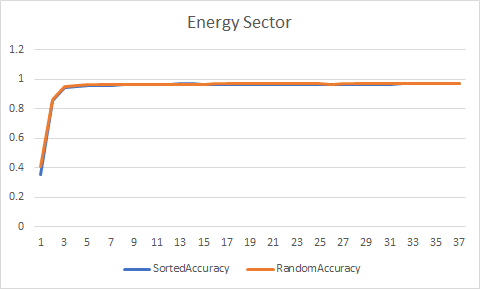}
		\end{minipage}%
	}%
\\
	\subfigure[Health Sector]{
		\begin{minipage}[t]{0.5\linewidth}
			\centering
			\includegraphics[width=2.2in]{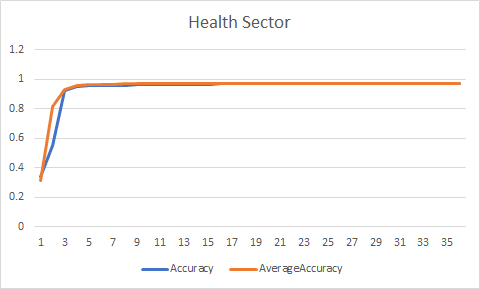}
		\end{minipage}
	}%
	\subfigure[Consumer Sector]{
		\begin{minipage}[t]{0.5\linewidth}
			\centering
			\includegraphics[width=2.2in]{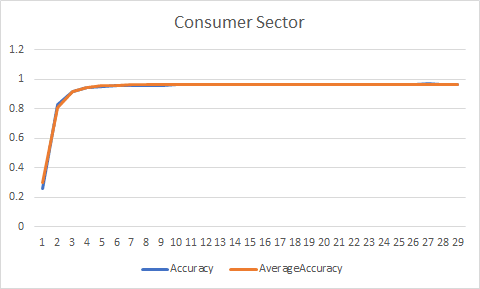}
		\end{minipage}
	}
	\caption{Feature number in Financial Ratio Detailed Mapping for Algorithm 2}
	\label{Feature number in Financial Ratio+Detailed Mapping for Algorithm 2}

\end{figure}

\begin{figure}[H]
	\centering

	\subfigure[Financial Sector]{
		\begin{minipage}{0.5\linewidth}
			\centering
			\includegraphics[width=2.2in]{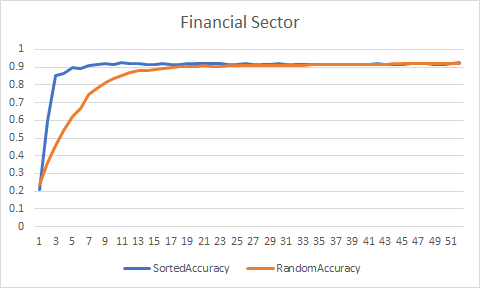}
		\end{minipage}%
	}%
	\subfigure[Health Sector]{
		\begin{minipage}{0.5\linewidth}
			\centering
			\includegraphics[width=2.2in]{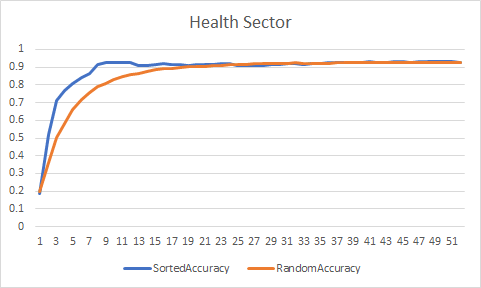}
		\end{minipage}%
	}%
\\
	\subfigure[Consumer Sector]{
		
			\centering
			\includegraphics[width=2.1in]{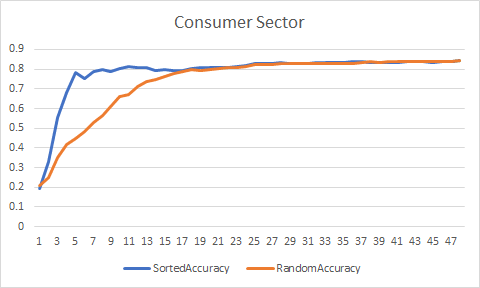}
	
	}%
		\caption{Feature Number  in Balance Sheet Detailed Mapping for Algorithm 2}
			\label{Accuracy in Balance Sheet +Detailed Mapping for Algorithm 2}
\end{figure}

\begin{figure}[H]
	\centering
	
	\subfigure[Financial Sector]{
		\begin{minipage}{0.5\linewidth}
			\centering
			\includegraphics[width=2.2in]{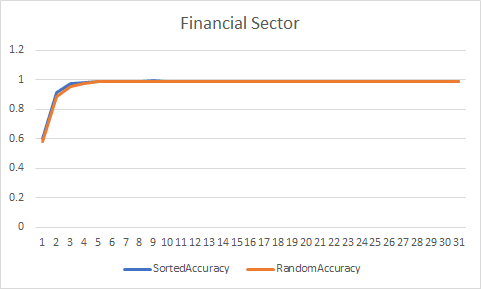}
		\end{minipage}%
	}%
	\subfigure[Energy Sector]{
		\begin{minipage}{0.5\linewidth}
			\centering
			\includegraphics[width=2.2in]{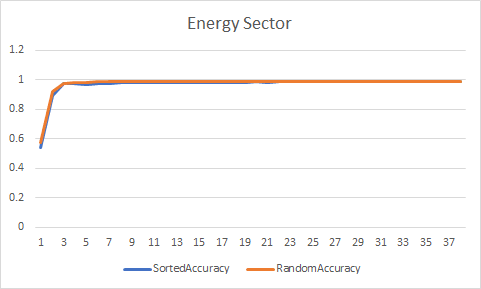}
		\end{minipage}%
	}%
\\
	\subfigure[Health Sector]{
		\begin{minipage}[t]{0.5\linewidth}
			\centering
			\includegraphics[width=2.2in]{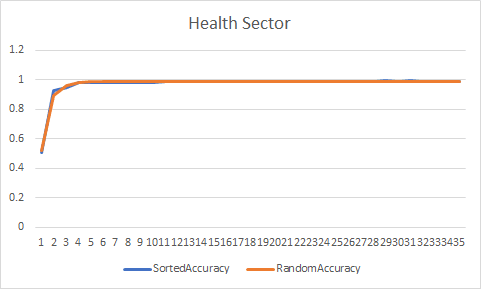}
		\end{minipage}
	}%
	\subfigure[Consumer Sector]{
		\begin{minipage}[t]{0.5\linewidth}
			\centering
			\includegraphics[width=2.2in]{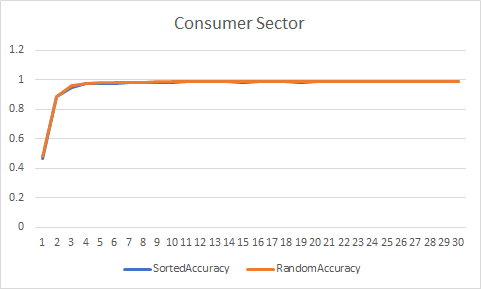}
		\end{minipage}
	}
		\caption{Feature number in Financial Ratio+Coarse Mapping for Algorithm 2}
		\label{Feature number in Financial Ratio+Coarse Mapping for Algorithm 2}

\end{figure}

\begin{figure}[H]
	\centering
		
	\subfigure[Financial Sector]{
		\begin{minipage}{0.5\linewidth}
			\centering
			\includegraphics[width=2.2in]{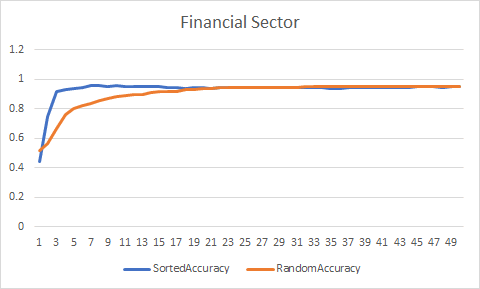}
		\end{minipage}%
	}%
	\subfigure[Health Sector]{
		\begin{minipage}{0.5\linewidth}
			\centering
			\includegraphics[width=2.2in]{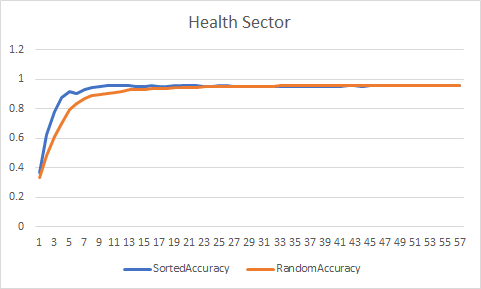}
		\end{minipage}%
	}%
	\\
	\subfigure[Health Sector]{
		\begin{minipage}{0.5\linewidth}
			\centering
			\includegraphics[width=2.2in]{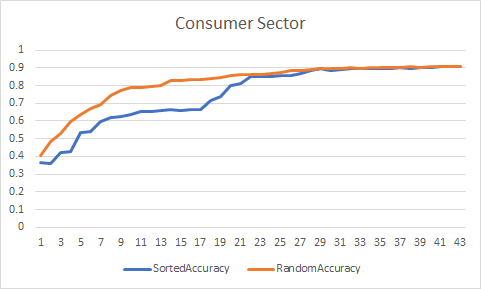}
		\end{minipage}%
	}%
\caption{Feature Number in  Balance Sheet Coarse Mapping for Algorithm 2}
		\label{Accuracy in Balance Sheet +Coarse Mapping for Algorithm 2}
\end{figure}

From Table \ref{The Number of original Features }, Table \ref{The Number of  Features selected by Chi-Square Test }, Table \ref{The Number of  Features for  Alogorithm 1 } , Table \ref{The Number of  Features selected for  Alogorithm 2 } , the number of feature sets of Algorithm 1 was equal to that of Chi-Suqare Test. The number of features selected by Algorithm 2 was less than that of Chi-square Test.  The features sets  were  sorted  in  decreasing order of components scores/factors scores and were carried out the iteration adding one feature in order in every time . We can obtain the minimum number of features when the accuracy curves became steady.To examine whether the sorted features better,the random order feature of every data  set also did this iteration and repeated 10 times(each time a feature set formed a new random feature set)  and the average accuracy of $n_{th}$ feature is recorded. The accuracy of comparison plot about sorted feature sets and random feature sets was displayed in Figure \ref{Feature number in Financial Ratio Detailed Mapping for Algorithm 1},  \ref{Feature Number in Balance Sheet Detailed Mapping for Algorithm 1},  \ref{Feature Number in Financial Ratio+Coarse Mapping for Algorithm 1},  \ref{Accuracy in Balance Sheet +Coarse Mapping for Algorithm 1},  \ref{Feature number in Financial Ratio+Detailed Mapping for Algorithm 2},  \ref{Accuracy in Balance Sheet +Detailed Mapping for Algorithm 2},  \ref{Feature number in Financial Ratio+Coarse Mapping for Algorithm 2},  \ref{Accuracy in Balance Sheet +Coarse Mapping for Algorithm 2}.

\begin{table}[H]
\centering
\caption{The Number of Features Shortilised by Algorithm 1}
\label{The Number of Features Shortilised by Algorithm 1}
\setlength\tabcolsep{3pt} 
\scriptsize
\begin{tabular}{ccccc  }
		
		\hline\hline\noalign{\smallskip}
		\quad & \tabincell{c}{Financial Raio\\Detailed Mapping} &\tabincell{c}{Financial Ratio\\Coarse Mapping}&\tabincell{c}{ Balance Sheet\\Detailed Mapping }&\tabincell{c}{ Balance Sheet\\Coarse Mapping } \\
		\noalign{\smallskip}\hline\noalign{\smallskip}
		\tabincell{c}{Financial Sector} &5  &12 & 54& 14\\
			\specialrule{0em}{1.5pt}{1.5pt}
			\tabincell{c}{Energy Sector} &10& 10&19 &19   \\
			\specialrule{0em}{1.5pt}{1.5pt}
      \tabincell{c}{Health Sector} &6 &8 & 66 &8 \\
      	\specialrule{0em}{1.5pt}{1.5pt}
      \tabincell{c}{  Consumer Sector }  & 33&8 &50  &40 \\
		
		\noalign{\smallskip}\hline\hline
	
	\end{tabular}
	\end{table}
\begin{table}[H]
\centering
\caption{The Number of Features Shortilised by Algorithm 2}
\label{The Number of Features Shortilised by Algorithm 2}
\scriptsize
\setlength\tabcolsep{3pt} 
\begin{tabular}{ccccc  }
		
		\hline\hline\noalign{\smallskip}
		\quad & \tabincell{c}{Financial Raio\\Detailed Mapping} &\tabincell{c}{Financial Ratio\\Coarse Mapping}&\tabincell{c}{ Balance Sheet\\Detailed Mapping }&\tabincell{c}{ Balance Sheet\\Coarse Mapping } \\
		\noalign{\smallskip}\hline\noalign{\smallskip}
		\tabincell{c}{Financial Sector} &5  &8 &21 &10 \\
			\specialrule{0em}{1.5pt}{1.5pt}
			\tabincell{c}{Energy Sector} &8&13 &NA&NA  \\
			\specialrule{0em}{1.5pt}{1.5pt}
      \tabincell{c}{Health Sector} & 9& 18& 13 &13 \\
      	\specialrule{0em}{1.5pt}{1.5pt}
      \tabincell{c}{  Consumer Sector }  &7 &13 & 32 & 37\\
		
		\noalign{\smallskip}\hline\hline
	
	\end{tabular}
	\end{table}

\begin{table}[H]
\centering
\caption{The Number of Random Features Shortilised by Algorithm 1}
\label{The Number of Random Features Shortilised by Algorithm 1}
\scriptsize
\setlength\tabcolsep{3pt} 
\begin{tabular}{ccccc  }
		
		\hline\hline\noalign{\smallskip}
		\quad & \tabincell{c}{Financial Raio\\Detailed Mapping} &\tabincell{c}{Financial Ratio\\Coarse Mapping}&\tabincell{c}{ Balance Sheet\\Detailed Mapping }&\tabincell{c}{ Balance Sheet\\Coarse Mapping } \\
		\noalign{\smallskip}\hline\noalign{\smallskip}
		\tabincell{c}{Financial Sector} & 11 &14 & 66& 19\\
			\specialrule{0em}{1.5pt}{1.5pt}
			\tabincell{c}{Energy Sector} &8 &7 &34 & 34 \\
			\specialrule{0em}{1.5pt}{1.5pt}
      \tabincell{c}{Health Sector} &8 &12 & 48 & 17\\
      	\specialrule{0em}{1.5pt}{1.5pt}
      \tabincell{c}{  Consumer Sector }  &38 &10 &57  &45 \\
		
		\noalign{\smallskip}\hline\hline
	
	\end{tabular}

	\end{table}
\begin{table}[H]
\centering
\caption{The Number of Random Features Shortilised by Algorithm 2}
\label{The Number of Random Features Shortilised by Algorithm 2}
\scriptsize
\setlength\tabcolsep{3pt} 
\begin{tabular}{ccccc  }
		
		\hline\hline\noalign{\smallskip}
		\quad & \tabincell{c}{Financial Raio\\Detailed Mapping} &\tabincell{c}{Financial Ratio\\Coarse Mapping}&\tabincell{c}{ Balance Sheet\\Detailed Mapping }&\tabincell{c}{ Balance Sheet\\Coarse Mapping } \\
		\noalign{\smallskip}\hline\noalign{\smallskip}
		\tabincell{c}{Financial Sector} & 6 & 9&47 &39\\
			\specialrule{0em}{1.5pt}{1.5pt}
			\tabincell{c}{Energy Sector} & 8& 13&NA &NA  \\
			\specialrule{0em}{1.5pt}{1.5pt}
      \tabincell{c}{Health Sector} &8 &17 & 49 &49 \\
      	\specialrule{0em}{1.5pt}{1.5pt}
      \tabincell{c}{  Consumer Sector }  &6 &11 &36  &34 \\
		
		\noalign{\smallskip}\hline\hline
	
	\end{tabular}

	\end{table}

   \begin{figure}[H]
	\centering
	
		\setlength\tabcolsep{3pt} 
	\subfigure[Financial Sector]{
		\begin{minipage}{0.5\linewidth}
			\centering
			\includegraphics[width=2.2in]{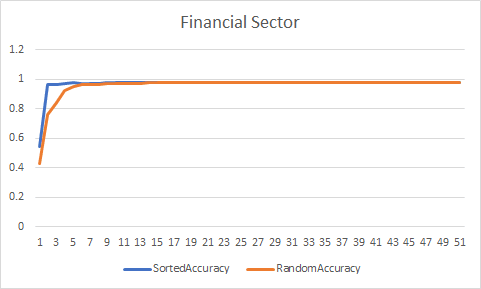}
		\end{minipage}%
	}%
	\subfigure[Energy Sector]{
		\begin{minipage}{0.5\linewidth}
			\centering
			\includegraphics[width=2.2in]{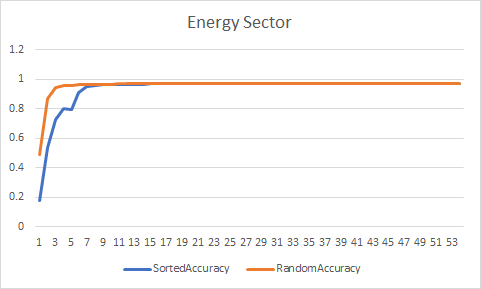}
		\end{minipage}%
	}%
\\
	\subfigure[Health Sector]{
		\begin{minipage}[t]{0.5\linewidth}
			\centering
			\includegraphics[width=2.2in]{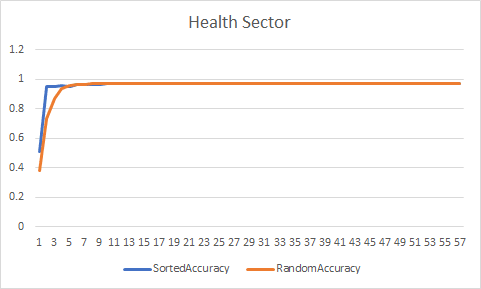}
		\end{minipage}
	}%
	\subfigure[Consumer Sector]{
		\begin{minipage}[t]{0.5\linewidth}
			\centering
			\includegraphics[width=2.2in]{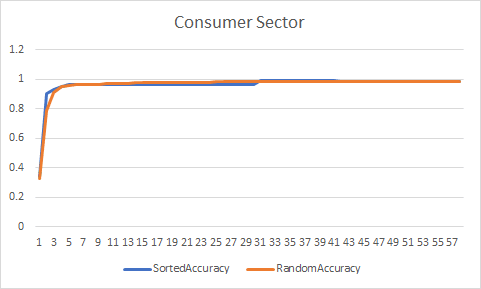}
		\end{minipage}
	}
		\caption{Feature number in Financial Ratio Detailed Mapping for Algorithm 1}
		\label{Feature number in Financial Ratio Detailed Mapping for Algorithm 1}

\end{figure}

 

From table \ref{The Number of Features Shortilised by Algorithm 1}, table \ref{The Number of Features Shortilised by Algorithm 2}, Algorithm 2 needed lesser number of feature  in comparison to Algorithm 1 for most of data sets to achieve almost the same accuracy that reduced significantly the time spent on data analyzing and  improved computational efficiency. In addition, the accuracy of Financial Ratio feature sets was always higher than that of Balance Sheet feature sets. From table \ref{The Number of Random Features Shortilised by Algorithm 1}, table \ref{The Number of Random Features Shortilised by Algorithm 2}, random order data sets need more number of features compared with sorted feature sets.  This confirmed sorting feature sets  before feature selecting was necessary   that improved significantly computational efficiency.
 \setcounter{table}{0}

\setcounter{figure}{0}

\section{Feature Classification}
Factor Analysis can  deal with data sets where there are a large number of observed variables that are thought to reflect a smaller number of latent variables.  
There are a large number of features that influence the credit rating change,Factor analysis can help to  summarize similar features together  to find out the latent factors that create a commonality    and  explain one public factor. In this Algorithm,  we can reduce significantly  the number of variables and obtain main factors that have effect on credit rating change to make people understand the reason of credit rating change better.
\\

The KMO value of Energy Sector in Balance sheet was 0.5 (less than 0.6)that indicated it was not adequate for Factor Analysis.Therefore, we do not perform Factor Analysis for Energy Sector in Balance Sheet.      
According to four data set for four sectors, we summarized the main factors that influenced the credit rating and displayed the results by summarizing the factors of   Financial Ratio and Balance Sheet for four sectors.
 \subsection{Financial Ratio  }
In this section, we summarized the Common factors in Financial Ratio for four sectors. There are seven factors that have influence on   credit rating, Debt/Asset and  Debt/capital  Factor, Earning and Profit  Factor, Cash Turnover Factor, Liquidity Factor, Stocks Value Factor,  Financing Factor, Enterprise Value Factor.

\subsubsection{Financial Ratio Factor 1}
The first factor is Debt/Asset and  Debt/capital  Factor. This factor includes Long-term Debt/Invested Capital, Total Debt/Capital, Total Debt/ \\Total Assets, Total Debt/Total Assets 1, Long-termDebt/Book Equity, Long-term Debt/Total Liabilities, Total Debt/Equity.\\

The Debt to Asset ratio indicates the company's  financial leverage and tells people the percentage of the company 's asset financed by debt, rather than equity.The higher the ratio, the higher the degree of leverage  and financial risk. The ratio can be used for investors to evaluate whether the company is reliable to meet its current debt obligation and pay a return on their investment. The debt to capital ratio gives analysts and investors a better understanding of the company's financial structure and whether the company is worth investing. Therefore, the higher the  Debt/Asset and  Debt/capital,the lower the credit rating.

 \begin{figure}[H]
	\centering
     
		\includegraphics[width=3.8in]{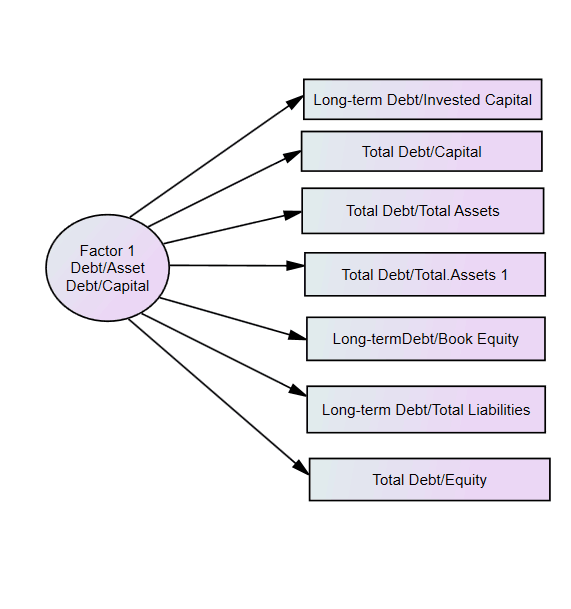}	
		\caption{Financial Ratio: Factor 1}
     \label{Financial Ratio: Factor 1}
	\end{figure}
\subsubsection{Financial Ratio Factor 2}
The second factor is Earning and Profit  Factor.
This factor includes Pre-tax Return on Total Earning Assets, Return on Capital Employed, After-tax Return on Average Common Equity, Return on Equity, Return on Assets, After-tax Return on
 Invested Capital, Operating Profit Margin
 After Depreciation, Net Profit Margin, Gross Profit /Total Assets, Profit /Before Depreciation Current Liabilities, Operating Profit Margin Before Depreciation.
 
Rerurns are the  earnings acquired before the interest is paid. Profit mainly included the gross profit margin, operating profit margin, and net profit margin. Profit is the money that company paid for all expense in production , sales ,taxes and so on. They are the indicator whether the company has enough funds to meet its current debt obligations and can run well in the long-run.

 \begin{figure}[H]
	\centering
     
		\includegraphics[width=3.5in]{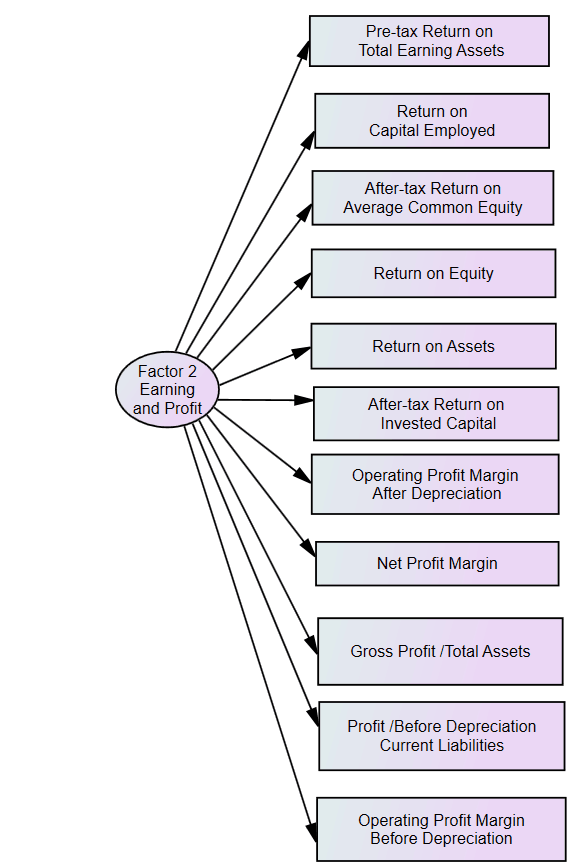}
		\caption{Financial Ratio Factor 2}
     \label{Financial Ratio: Factor 2}
	\end{figure}

\subsubsection{Financial Ratio Factor 3}
The third factor is  Cash Turnover  Factor.
This factor includes Asset Turnover, Receivables Turnover, Inventory /Current.Assets, Payables /\\Turnover.
Turnover is an accounting concept that calculates the speed at which a  company operates. Turnover ratio calculates the speed at which a company collects cash from accounts receivable and inventory investments. Analysts and investors use these ratios to determine whether the company is seen as a good investment.

 \begin{figure}[H]
	\centering
     
		\includegraphics[width=3.5in]{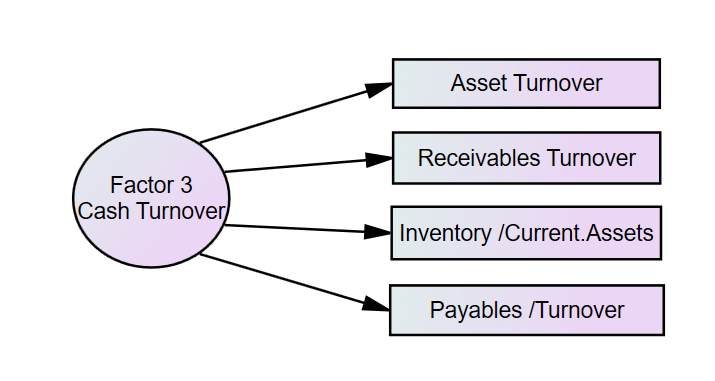}
		\caption{Financial Ratio Factor 3}
     \label{Financial Ratio: Factor 3}
	\end{figure}

\subsubsection{Financial Ratio Factor 4}
This factor is Liquidity  Factor.
This factor includes  Cash Ratio, Cash Balance/Total Liabilities,Quick Ratio (Acid Test), Current Ratio. 
The cash ratio is a measure of company  liquidity, especially the ratio of company cash and cash equivalents to its current liabilities.The cash ratio is a liquidity indicator of the ability of companies to use only cash and cash equivalents to cover their short-term debt.

 \begin{figure}[H]
	\centering
    
		\includegraphics[width=4in]{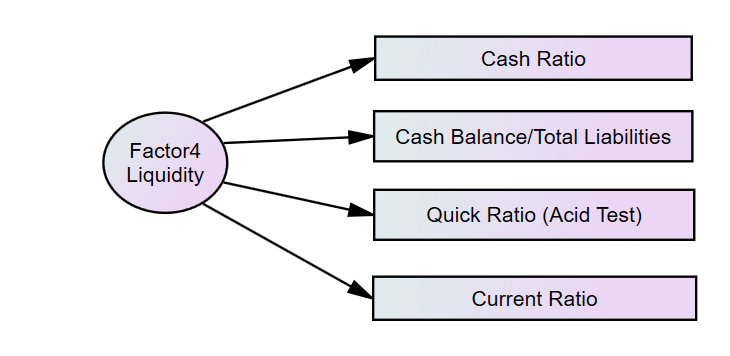}
		
		 \caption{Financial Ratio: Factor 4}
     \label{Financial Ratio: Factor 4}
	\end{figure}

\subsubsection{Financial Ratio Factor 5}
This factor is Stocks Value  Factor.
This factor includes P/E(Diluted,Excl,EI), P/E(Diluted,Incl,EI).
The P / E ratio is the valuation ratio of its current share price relative to its earnings per share.Investors and analysts use the P / E ratio to determine the relative value of the company's shares for comparison with the company. 
 \begin{figure}[H]
	\centering
    
		\includegraphics[width=3in]{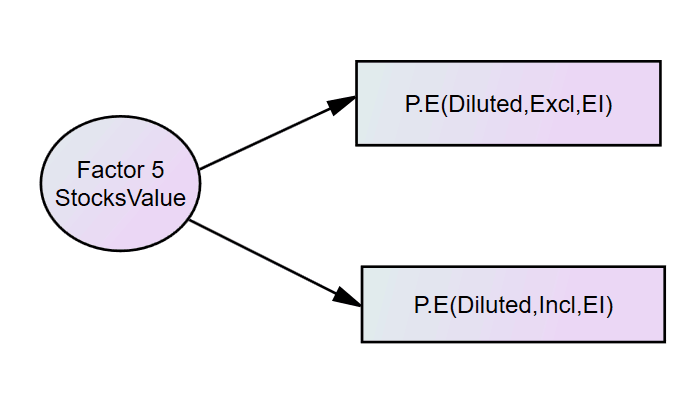}	
		
		 \caption{Financial Ratio Factor 5}
     \label{Financial Ratio: Factor 5}
	\end{figure}

\subsubsection{Financial Ratio Factor 6}
This factor is Financing  Factor.
This factor includes Short-Term Debt/Total Debt, Long-term Debt/
Total Liabilities, Current Liabilities /Total Liabilities.
Total liabilities are the sum of individual or company debts.They are generally divided into three categories: short-term liabilities, long-term liabilities and other liabilities.Short-Term Debt/Total Debt  indicates the extent of a company's reliance on short-term financing.Long-term Debt/ Total Liabilities shows the degree of a company's reliance on long-term financing.The larger the proportion of debt, the higher the financial leverage of the company, the less stable the company has its own capital, and the need to consider paying off the debt at any time, which will require higher cash flow of the company, so that the operating risk of the company will increase.

 \begin{figure}[H]
	\centering
    
		\includegraphics[width=3in]{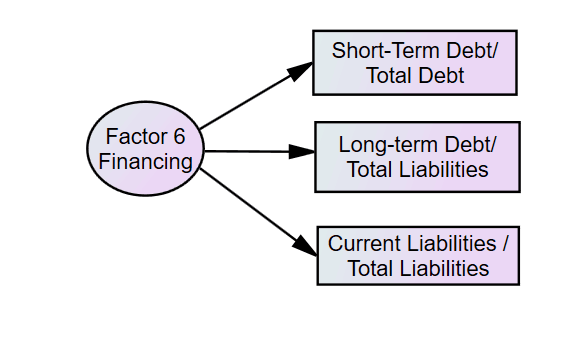}	
		
		 \caption{Financial Ratio Factor 6}
     \label{Financial Ratio: Factor 6}
	\end{figure}
	
\subsubsection{Financial Ratio Factor 7}
This factor is Enterprise Value  Factor.
This factor includes Enterprise Value Multiple, Total Debt /EBITDA.
Enterprise multiple is a ratio used to evaluate the value of a company. Enterprise multiples are calculated by EBITDA according to the value of the enterprise. EBITDA is an abbreviation for earnings before interest, tax, depreciation and amortization. Enterprise multiples is used to determine whether a company is undervalue or overvalue for investors.The lower ratio, the company is more likely to be undervalue.The higher ratio, the company is more likely to be overvalue.

 \begin{figure}[H]
	\centering
     
		\includegraphics[width=3.5in]{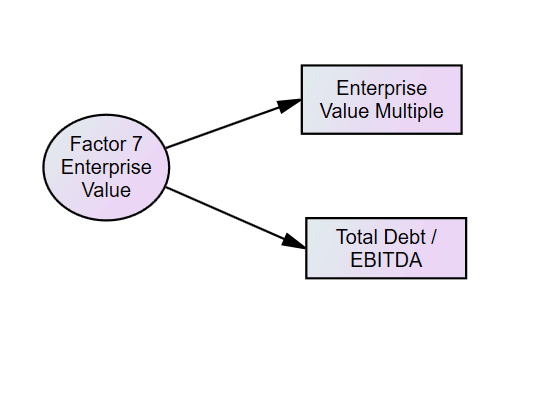}	
		
		\caption{Financial Ratio Factor 7}
     \label{Financial Ratio: Factor 7}
	\end{figure}
\subsection{Balance Sheet}

In this section, we summarized the Common factors in Balance Sheet  for four sectors. There are ten factors influencing the credit rating, Asset and liabilitied Factor, Income Factor, Implied Option Factor, Post Retirement Adjustment Factor, Pension Adjustment Factor, Earing Per Share Factor, Discontinued Operations Factor, Preferred  Divifends Factor, Non-Controlling  Interests  Factor, and In  Process  R$\&$D  Factor.
\subsubsection{Balance Sheet Factor 1}
The first  factor is Asset and Liabilities  Factor.
This factor includes Current Assets -Other - Total, Current Liabilities -Other - Total.
Current assets can reflect the operation of the company  that they can be used to support daily business operation and to pay for operating expenses.The current liabilities refer to the debts or liabilities that the Company has due within one year or within the normal operating period. In addition, current liabilities are settled through the use of current assets, such as cash.Therefore, this factor represent the state of operation for a company.

 \begin{figure}[H]
	\centering
    
		\includegraphics[width=3.5in]{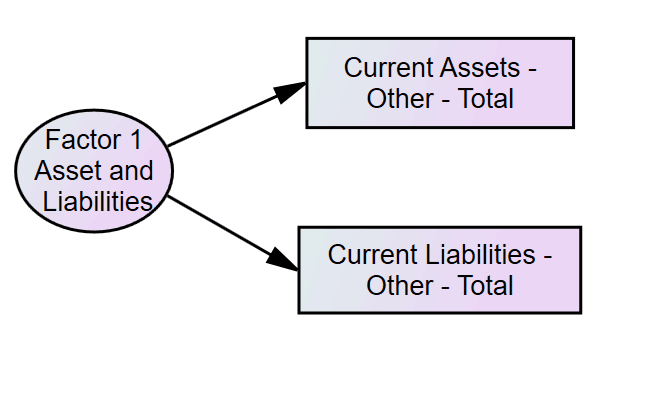}
		 \caption{Balance Sheet Factor 1}
     \label{Balance Sheet: Factor 1}
	\end{figure}

\subsubsection{Balance Sheet Factor 2}
The second  factor is Income Factor.
This factor includes Comp Inc - Beginning Net Income, Pretax Income, Comprehensive Income-Parent, Special Items, Income Before Extra Items -Adj for Common Stock Equivalents - 12MM.Income is the company's remaining revenues after paying all opeating expenses and taxes.The higher the income, the probably better operating state.

 \begin{figure}[H]
	\centering
     
		\includegraphics[width=4in]{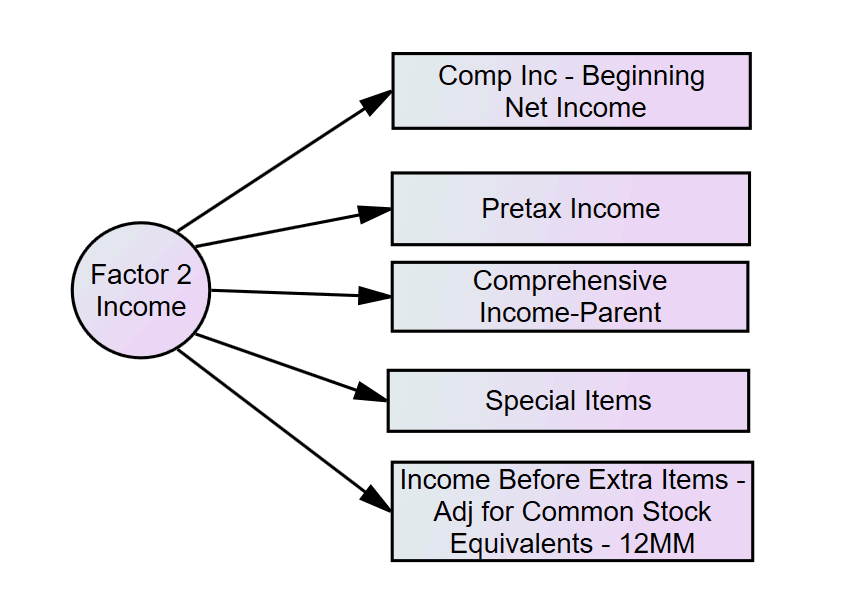}	\caption{Balance Sheet Factor 2}
     \label{Balance Sheet: Factor 2}
	\end{figure}

\subsubsection{Balance Sheet Factor 3}
The third factor is Implied
Option Factor.
This factor includes Implied Option EPS Diluted, Implied Option EPS Diluted Preliminary, Implied Option EPS Diluted 12MM, Implied Option Expense, Implied Option Expense Preliminary, Implied Option
Expense - 12mm.
The implied option mainly refers to the pre-redemption option, which is the risk of the interest rate that the borrower can repay the risk in advance.The existence of implied options is one of the important reasons for financial risk.

 \begin{figure}[H]
	\centering
    
		\includegraphics[width=4in]{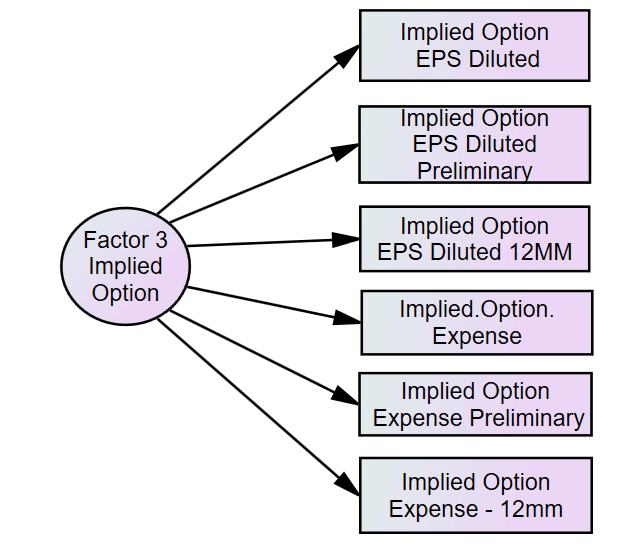}	 \caption{Balance Sheet Factor 3}
     \label{Balance Sheet: Factor 3}
	\end{figure}

\subsubsection{Balance Sheet Factor 4}
The third factor is Post Retirement
Adjustment Factor.
This factor includes Core Post Retirement
Adjustment, Core Post Retirement Adjustment 12MM, Core Post Retirement Adjustment Diluted EPS Effect 12MM, Core Post Retirement Adjustment Diluted EPS Effect, Core Post Retirement Adjustment 12MM Diluted EPS Effect Preliminary.Retirement Adjustment  will have effect on employees, which will influence operating of the company.

 \begin{figure}[H]
	\centering
    
		\includegraphics[width=4in]{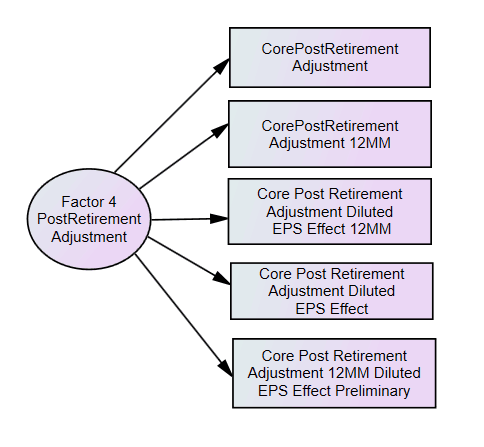}	 \caption{Balance Sheet Factor 4}
     \label{Balance Sheet: Factor 4}
	\end{figure}
\subsubsection{Balance Sheet Factor 5}
This factor is  Pension Adjustment Factor.This factor includes Core Pension Interest Adjustment After-tax Preliminary, Core Pension Adjustment Preliminary, Core Pension Adjustment, Core Pension Adjustment Diluted EPS Effect 12MM, Core Pension Adjustment Diluted EPS Effect.The Pension Adjustment is a combination of all employee and employer pension credits for the year. The higher pension, the company have more funds to operate.

 \begin{figure}[H]
	\centering
    
		\includegraphics[width=4in]{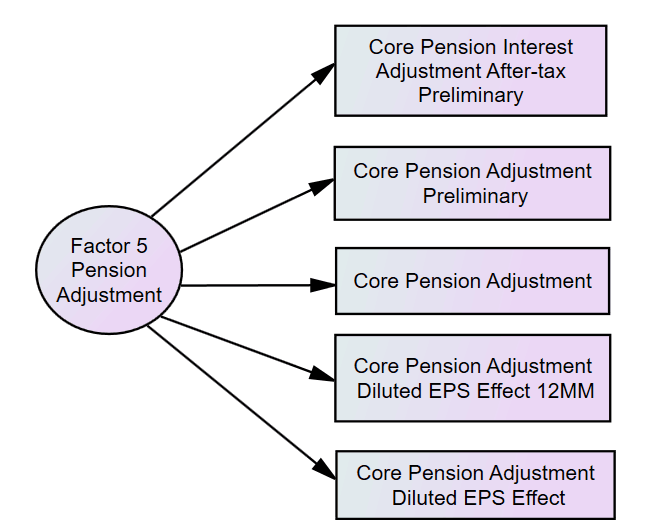}	 \caption{Balance Sheet Factor 5}
     \label{Balance Sheet: Factor 5}
	\end{figure}

\subsubsection{Balance Sheet Factor 6}
This factor is  Earnings Per Share Factor.This factor includes Earnings Per Share (Diluted)-Excluding Extraordinary Items- 12 Months Moving, Earnings Per Share (Diluted)-Including Extraordinary Items.1, Core Pension w/o Interest Adjustment Diluted EPS Effect Preliminary. Earnings Per Share is the share of the company's profits divided by its common stock. It indicates the profitability of a company. The higher the  Earnings Per Share , the more profit a company obtains.

 \begin{figure}[H]
	\centering
    
		\includegraphics[width=4in]{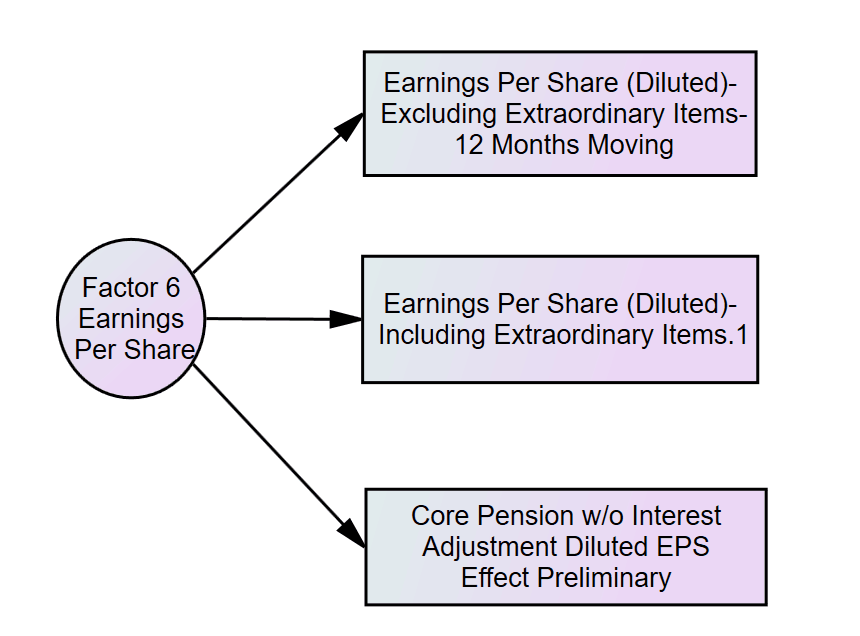}	 \caption{Balance Sheet Factor 6}
     \label{Balance Sheet: Factor 6}
	\end{figure}
\subsubsection{Balance Sheet Factor 7}
This factor is Discontinued Operations Factor. This factor includes Discontinued Operations, Extraordinary Items and Discontinued Operations\\Discontinued operations  is shown separately in the income statement, and investors can clearly know the  profits and cash flows from the discontinued activities, especially when the company is merged, and it is better to analyze which assets are stripped or folded to understand how a company will make money in the future.

 \begin{figure}[H]
	\centering
    
		\includegraphics[width=4in]{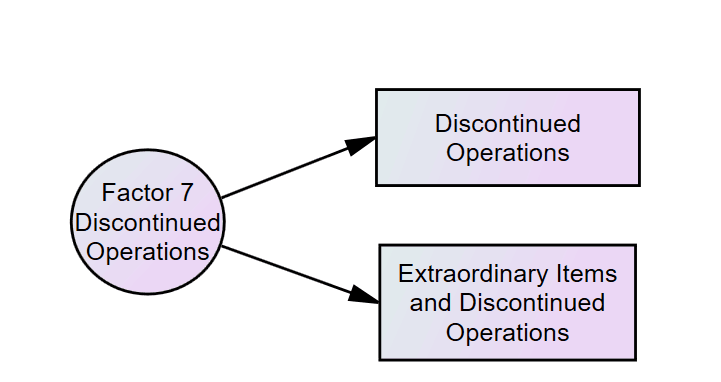}	 \caption{Balance Sheet Factor 7}
     \label{Balance Sheet: Factor 7}
	\end{figure}

\subsubsection{Balance Sheet Factor 8}

This factor is Preferred Dividends Factor. This factor includes Nonred Pfd Shares Outs (000) - Quarterly, Preferred/Preference
Stock (Capital)-Total, Dividends-Preferred/Preference.A preferred dividend refers to the dividends accrued and paid by the company's preferred shares.The higher the company's earnings per share, the higher its profitability.

 \begin{figure}[H]
	\centering
    
		\includegraphics[width=3in]{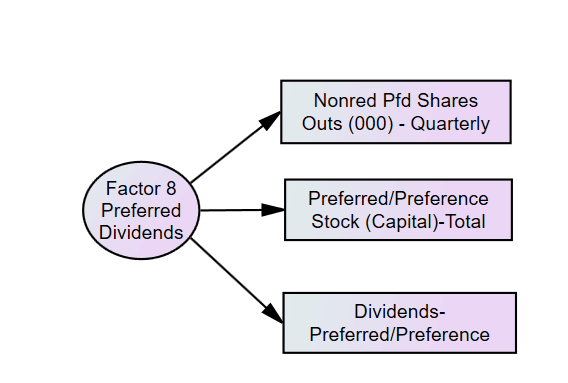}	 \caption{Balance Sheet Factor 8}
     \label{Balance Sheet: Factor 8}
	\end{figure}
\subsubsection{Balance Sheet Factor 9}
This factor is Non-Controlling Interests Factor. This factor includes
Non-Controlling Interest - Redeemable - Balance Sheet, Non-Controlling Interests - Total - Balance Sheet.
Non-Controlling  Interests is not owned by the parent company, which owns a controlling stake (more than 50 per cent but less than 100 per cent) and merges the financial performance of the subsidiary into its own company.For investors, it is important for companies to provide transparency of non-controlled benefits, because this will enable them to better understand the impact of non-controlled benefits on the financial position, financial performance and cash flow of the company, and thus also understand the risks the company faced. 
 \begin{figure}[H]
	\centering
    
		\includegraphics[width=4in]{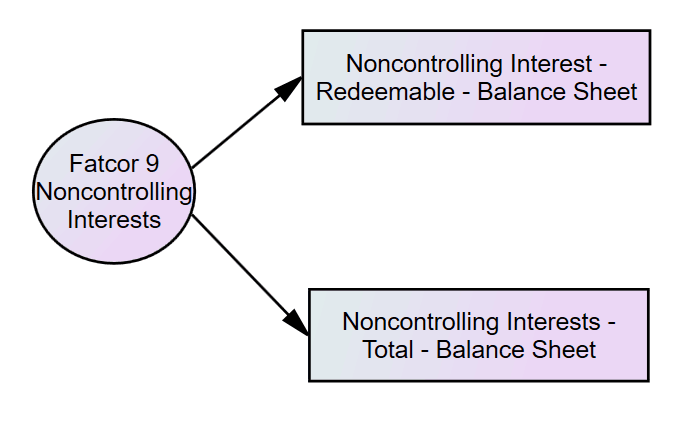}	 \caption{Balance Sheet Factor 9}
     \label{Balance Sheet: Factor 9}
	\end{figure}

\subsubsection{Balance Sheet Factor 10}
This factor is In Process R\&D Factor. This factor includes In Process R\&D, In Process R\&D
Expense After-tax.A company typically generates research and development costs as it seeks and creates new products or services.It is important for the company to maintain the forefront of innovation and invest heavily in product research and development. With the  development of technology, these efforts enable the company to diversify and find new growth opportunities.

\begin{figure}[H]
	\centering
     
		\includegraphics[width=3.6in]{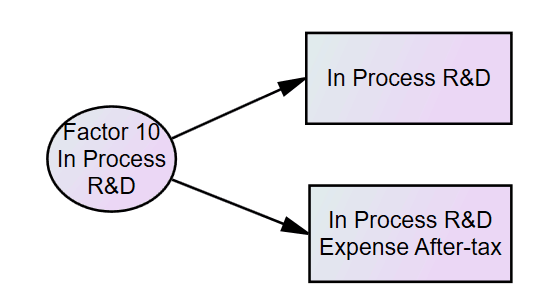}	
		\centering	
		\caption{Balance Sheet Factor 10}
     \label{Balance Sheet: Factor 10}
	\end{figure}

\section{Conclusion}
In this paper, we selected the important features that had a significant impact on the corporate credit via  Principle Component Analysis and Factor Analysis, and compared which one can better select the essential features to explain the reason for credit rating change. We optimized the sorting method of principal component analysis and factor analysis. For Principle Component Analysis, we added the   the squares of loadings of every component of each feature instead of  adding absolute values to avoid the influence of extreme values. As for Factor Analysis,  according to the principle of the
first factor priority, we sorted each factor’s features per the absolute value of loadings within
the first factor and then the second factor and downwards. Then, the  results showed that Factors Analysis required fewer
features compared to the Principal Component Analysis to actuate the same accuracy of
above 95\%. Finally, we concluded  dominant factors on credit rating, in the data set of Financial Ration, seven factors influencing the corporate: Debt/Asset and Debt/capital Factor, Earningand Profit Factor, Cash Turnover Factor, Liquidity Factor, Stocks Value Factor, Financing Factor,Enterprise Value Factor; in Balance Sheet, ten factors affecting the credit rating: Asset and liabilitied Factor, Income Factor, Implied Option Factor, Post Retirement Adjustment Factor, Pension Adjustment Factor, Earing Per Share Factor, Discontinued Operations Factor, Preferred  Divifends Factor, Non-Controlling  Interests  Factor, and In  Process  R$\&$D  Factor.

 \setcounter{table}{0}

\setcounter{figure}{0}

\newpage
 
   \addappheadtotoc
   \appendixpage

 \appendix
  \renewcommand{\appendixname}{Appendix~\Alph{section}}
  \section{Top 10 features selected by Algorithm 1}
  \subsection{Financial Sector}
  \subsubsection{Financial Sector:Data1: Financial Ratio + Detailed Mapping}
 Receivables Turnover\\
Total Debt/Equity\\
Payables Turnover\\
Capitalization Ratio\\
Operating Profit Margin After Depreciation\\
Total Debt/Total Assets 1\\
Long-term Debt/Invested Capital\\
Research and Development/Sales\\
Total Liabilities/Total Tangible.Assets\\
Book/Market\\
 \subsubsection{Financial Sector:Data2:Balance Sheet + Detailed Mapping}
Noncontrolling Interest - Redeemable - Balance Sheet\\
Deferred Taxes and Investment Tax Credit\\
Common Shares Used to Calculate Earnings Per Share - 12 Months Moving\\
Cost of Goods Sold\\
Accounting Changes - Cumulative Effect\\
Stock Compensation Expense\\
Preferred ESOP Obligation - Non-Redeemable\\
Property Plant and Equipment - Total (Net)\\
Core Pension Adjustment Diluted EPS Effect\\
Extraordinary Items\\

 \subsubsection{Financial Sector:Data3 :Financial Ratio + Coarse Mapping}
 Receivables Turnover\\
Total Debt/Equity\\
Long-term Debt/Invested Capital\\
Research and Development/Sales\\
Capitalization Ratio\\
Payables Turnover\\
P/E (Diluted, Excl. EI)\\
Short-Term Debt/Total Debt\\
Total Debt/Total Assets.1\\
Total Debt/Capital\\
 \subsubsection{Financial Sector:Data 4 :Balance Sheet + Coarse Mapping}
Core Post Retirement Adjustment Diluted EPS Effect 12MM\\
Interest and Related Expense-Total\\
Implied Option EPS Diluted Preliminary\\
Preferred ESOP Obligation-Non.Redeemable\\
Implied Option Expense Preliminary\\
Pretax Income\\
Implied Option EPS Diluted\\
Current Assets-Other-Total\\
Cash and Short Term Investments\\
Carrying Value\\
 \subsection{Energy Sector}
   \subsubsection{Energy Sector :Data1: Financial Ratio + Detailed Mapping}
Price/Cash flow\\
Receivables Turnover\\
P.E(Diluted,Excl,EI)\\
Research and Development/Sales\\
Avertising Expenses/Sales\\
Cash Flow Margin\\
Total Liabilities /Total Tangible Assets\\
Dividend Payout Ratio\\
Short-Term Debt/Total Debt\\
Common Equity/Invested Capital\\

  \subsubsection{Energy Sector :Data2:Balance Sheet + Detailed Mapping}
  Non-Operating Income (Expense) - Total\\
Other.Long.term.Assets\\
Gain Loss on Sale-Core Earnings Adjusted After-tax 12MM\\
Non-Current Assets-Total\\
Comp Inc-Derivative Gains Losses\\
Derivative Liabilities Long-Term\\
Pension Core Adjustment-12mm\\
Goodwill(net)\\
Current Assets-Other-Total\\
Income Before Extra-Items-Adjfor Common Stock Equivalents-12MM\\

\subsubsection{Energy Sector : Data3 :Financial Ratio + Coarse Mapping}
P/E (Diluted, Excl, EI)\\
Book/Market\\
Operating.CF.Current.Liabilities\\
P/E (Diluted, Incl. EI)\\
Short-Term Debt/Total Debt\\
Price/Book\\
Return.on.Assets\\
Return.on.Equity\\
Net Profit Margin\\
Inventory/Current Assets\\

 \subsubsection{ Energy Sector :Data 4 :Balance Sheet + Coarse Mapping}
 Core Post Retirement Adjustment 12MM Diluted EPS Effect Preliminary\\
Interest and Related Expense-Total\\
Implied Option EPS Diluted Preliminary\\
Preferred ESOP Obligation-Non Redeemable\\\
Implied Option Expense Preliminary\\
Pretax Income\\
Implied Option EPS Diluted\\
Current Assets-Other-Total\\
Cash and Short-Term Investments\\
Carrying Value\\

\subsection{Health Sector}
   \subsubsection{Health  Sector :Data1: Financial Ratio + Detailed Mapping}
   
   Pre-tax Return on Total Earning Assets\\
Free Cash Flow Operating Cash Flow\\
After-tax Return on Average Common Equity\\
Inventory/Current Assets\\
Common Equity /Invested Capital\\
Operating CF/CurrentLiabilities\\
Sales/Stockholders Equity\
Sales/Working Capital\\
Price/Sales\\
ReturnonAssets\\

  \subsubsection{Health Sector :Data2:Balance Sheet + Detailed Mapping}

  Core Post Retirement Adjustment 12MM Diluted EPS Effect Preliminary\\
In Process RD Expense Aftertax\\
Core Post Retirement Adjustment Diluted EPS Effect 12MM\\
Debtin Current Liabilities\\
Current Liabilities-Other-Total\\
Earnings PerShare(Diluted)-Excluding Extraordinary items\\
SPCore Earnings EPS Diluted Preliminary\\
Interestand Related Expense Total\\
Total Long-term Investments\\
Core Post Retirement Adjustment Preliminary\\

\subsubsection{Health  Sector : Data3 :Financial Ratio + Coarse Mapping}
P/E (Diluted, Excl, EI)
Cash Flow/Total Debt\\
PretaxreturnonNet/OperatingAssets\\
GrossProfit/TotalAssets\\
CashBalance/TotalLiabilities\\
TotalDebt/TotalAssets\\
Effective Tax Rate\\
ProfitBeforeDepreciation/CurrentLiabilities\\
Price/Book\\
Total Liabilities/Total Tangible Assets\\

 \subsubsection{Health Sector :Data 4 :Balance Sheet + Coarse Mapping} 
Preferred ESOP Obligation - Non-Redeemable\\
In Process R\&D Expense After-tax\\
Depreciation, Depletion and Amortization (Accumulated)\\
Treasury Stock - Total (All Capital)\\
Non-controlling Interests - Total- Balance Sheet\\
Non-Current Assets - Total\\
Core Post Retirement Adjustment Diluted EPS Effect 12MM \\
Core Post Retirement Adjustment 12MM Diluted EPS Effect Preliminary\\
Implied Option Expense\\
Core Post Retirement Adjustment Preliminary\\

\subsection{Consumer Sector}
   \subsubsection{Consumer  Sector :Data1: Financial Ratio + Detailed Mapping}
   P/E (Diluted,Incl,EI)
Common.Equity/Invested.Capital
Long.term.Debt/Invested.Capital
Pre-tax.Return on Total Earning Assets
Net Profit Margin
Gross Profit Margin
Effective Tax Rate
Current Liabilities/Total Liabilities
Total Debt/Total Assets
Profit Before Depreciation/Current Liabilities

  \subsubsection{Consumer Sector :Data2:Balance Sheet + Detailed Mapping}
  Operating Income Before Depreciation-Quarterly\\
Non Operating Income(Expense)-Total\\
Core Pension Adjustment\\
Core Post Retirement Adjustment 12MM Diluted EPS Effect Preliminary\\
Common Shares Used to Calculate Earnings Per Share - 12 Months Moving\\
Dilution Adjustment\\
Accum Other Comp Inc - Derivatives Unrealized Gain/Loss\\
Core Post Retirement Adjustment Preliminary\\
Accounting Changes-Cumulative Effect\\

\subsubsection{Consumer Sector : Data3 :Financial Ratio + Coarse Mapping}
Trailing P/E to Growth (PEG) ratio\\
Common Equity/Invested Capital\\
Pre-tax Return on Total Earning Assets\\
Return on Assets\\
Gross Profit Margin\\
Long-term Debt/Invested Capital\\
Net Profit Margin\\
Effective Tax Rate\\
Long-term Debt/Total Liabilities\\
Short Term Debt/Total Debt\\

\subsubsection{Consumer  Sector :Data 4 :Balance Sheet + Coarse Mapping}

Operating.Income.Before.Depreciation-Quarterly\\
Non-Operating Income(Expense)-Total\\
Core Pension Adjustment\\
Core Post Retirement Adjustment 12MM Diluted EPS Effect Preliminary\\
Common Shares Used to Calculate Earnings Per Share-12.Months.Moving\\
Dilution Adjustment\\
Accum Other Comp Inc - Derivatives Unrealized Gain/Loss\\
Core Post Retirement Adjustment.Preliminary\\
Accounting Changes-Cumulative.Effect\\

  \section{Top 10 features selected by Algorithm 2}
  \subsection{Financial Sector}
   \subsubsection{Financial Sector:Data1: Financial Ratio + Detailed Mapping}
 Total Debt/Total.Assets\\
Total Debt/EBITDA\\
Long -term.Debt/Total Liabilities\\
Cash Flow/Total.Debt\\
Long-term Debt/Book Equity\\
Total Debt/Capital\\
Total Debt/Equity\\
Asset Turnover\\
Receivables Turnover\\
Payables Turnover\\

  \subsubsection{Financial Sector:Data2:Balance Sheet + Detailed Mapping}
 Account Payable Creditors-Trade\\
Long-Term Debt-Total\\
Capital Surplus Share Premium Reserve\\
Cash and Short-Term Investments\\
Goodwill(net)\\
Non-Operating Income (Expense) - Total\\
Common Shares Used to Calculate Earnings Per Share - 12 Months Moving\\
Preferred/Preference Stock - Nonredeemable\\
Sales Turnover (Net)\\
Depreciation and Amortization-Total\\

\subsubsection{ Financial Sector:Data3 :Financial Ratio + Coarse Mapping}
Total Debt /Total Assets 1\\
Asset Turnover\\
Return on Assets\\
Total Debt/Equity\\
Cash Flow/Total.Debt\\
Long-term Debt/Total Liabilities\\
Total Debt/Capital\\
Receivables Turnover\\
Gross.Profit/Total Assets\\
Capitalization Ratio\\

 \subsubsection{ Financial Sector:Data 4 :Balance Sheet + Coarse Mapping}
Account Payable/Creditors-Trade\\
Long-Term Debt-Total\\
Capital Surplus/Share Premium Reserve\\
Cash and Short-Term Investments\\
Goodwill(net)\\
Non-Operating Income (Expense)-Total\\
Common Shares Used to Calculate Earnings Per Share-12 Months Moving\\
Sales Turnover (Net)\\
Depreciation and Amortization -Total\\
Retained Earnings\\

\subsection{Energy Sector}
   \subsubsection{Energy Sector :Data1: Financial Ratio + Detailed Mapping}
Pre-tax Return on Total Earning Assets\\
Return on Capital Employed\\
After-tax Return on Average Common Equity\\
Return on Equity\\
Return on Assets\\
After-tax Return on Invested Capital\\
Operating Profit.Margin.After.Depreciation\\
Net Profit Margin\\
Gross Profit/Total Assets\\
Profit Before Depreciation/Current Liabilities\\

\subsubsection{Energy Sector : Data3 :Financial Ratio + Coarse Mapping}
 Pre-tax Return on Total Earning Assets\\
Return on Capital Employed\\
After-tax Return on Average Common Equity\\
Return on Equity\\
Return on Assets\\
After-tax Return on Invested Capital\\
Operating Profit Margin After Depreciation\\
Net Profit Margin\\
Gross Profit/Total Assets\\
Profit Before Depreciation /Current iabilities\\

\subsection{Health Sector}
   \subsubsection{Health  Sector :Data1: Financial Ratio + Detailed Mapping}
  Long-term Debt/Invested Capital\\
Common Equity/Invested Capital\\
Total Debt/Invested Capital\\
Total Debt/Total Assets\\
Total Debt/Capital\\
Long-term Debt/Book Equity\\
Total Debt/Total Assets1\\
Long-term Debt/Total Liabilities\\
Return on Equity\\
After-tax Return onI nvested Capital\\

  \subsubsection{Health Sector :Data2:Balance Sheet + Detailed Mapping}
  Core Post Retirement Adjustment 12MM Diluted EPS Effect Preliminary\\
In Process R\&D Expense After-tax\\
Core Post Retirement Adjustment Diluted EPS Effect 12MM\\
Debtin Current Liabilities\\
Current Liabilities-Other-Total\\
Earnings Per Share(Diluted)-Excluding Extraordinaryitems\\
SP Core Earnings EPS Diluted Preliminary\\
Interestand Related Expense Total\\
Total Long-term Investments\\
Core Post Retirement Adjustment Preliminary\\

\subsubsection{Health  Sector : Data3 :Financial Ratio + Coarse Mapping}
 Long-term Debt/Invested Capital\\
Total Debt/Invested Capital\\
Common EquityInvested Capital\\
Total Debt/Total Assets\\
Total Debt/Capital\\
Long-term Debt Book Equity\\
TotalDebt/Total Assets1\\
Long-termDebt/Total Liabilities\\
Return on Equity\\
After-tax Return on Invested Capital\\

 \subsubsection{Health Sector :Data 4 :Balance Sheet + Coarse Mapping} 
Non-Current Assets-Total\\
Current Liabilities-Other-Total\\
Long-Term Liabilities-Total\\
Current Assets-Total\\
Depreciation and Amortization-Total\\
Operating Income After Depreciation Quarterly\\
Common Shares Used to Calculate Earnings Per Share-12 Months Moving\\
Current Assets-Other-Total\\
Property Plant and Equipment-Total(Net)\\
Research and Development Expense\\

\subsection{Consumer Sector}
   \subsubsection{Consumer  Sector :Data1: Financial Ratio + Detailed
   Mapping}
   Pre-tax Return on Total Earning Assets\\
Return on Assets\\
Profit Before Depreciation Current Liabilities\\
Accruals Average Assets\\
Return on Capital Employed\\
Total Debt/Total Assets\\
Total Debt/Total Assets 1\\
Total.Debt/Capital\\
Cash Ratio\\
Quick Ratio(Acid)Test \\

  \subsubsection{Consumer Sector :Data2:Balance Sheet + Detailed Mapping}
  Interest and Related Expense-Total\\
Cost of Goods Sold\\
Depreciation and Amortization-Total\\
Long-Term Debt-Total\\
Property Plant and Equipment - Total (Net)\\
Operating Income Before Depreciation-Quarterly\\
Liabilities-Other\\
Current Liabilities-Other-Total\\
Operating Income After Depreciation-Quarterly\\
Assets - Other - Total\\

\subsubsection{Consumer Sector : Data3 :Financial Ratio + Coarse Mapping}

Pre-tax Return on Total Earning Assets\\
Return on Assets\\
Profit Before Depreciation Current Liabilities\\
Accruals Average Assets\\
Return on Capital/Employed\\
Total Debt/Total Assets\\
Total Debt/Total Assets 1\\
Total Debt/Capital\\
Cash Ratio\\
Quick Ratio(Acid)Test\\
\subsubsection{Consumer  Sector :Data 4 :Balance Sheet + Coarse Mapping}
Special Items\\
Pretax Income\\
Comp Inc(Beginning)Net Income\\
Redeem Pfd Shares Outs (000)\\
Earnings Per Share(Diluted)-Including Extraordinary Items 1\\
Earnings Per Share(Diluted)from Operations\\
Income Before Extra Items-Adj for Common Stock Equivalents-12MM\\
Pension Core Adjustment-12mm\\
Core Pension Adjustment Diluted EPS Effect 12MM\\
Core Pension Adjustment Preliminary\\

\end{document}